\documentclass[aps, notitlepage,nofootinbib,twocolumn,longbibliography,superscriptaddress]{revtex4-2}

\newcommand{\be}{\begin{equation}}
\newcommand{\ee}{\end{equation}}

\usepackage[colorlinks=true,citecolor=blue,linkcolor=blue]{hyperref}
\usepackage{graphicx}
\usepackage{dcolumn}
\usepackage{bm}
\usepackage{color}
\usepackage{tabularx}
\usepackage{comment}
\usepackage{float}
\usepackage{amsmath}
\usepackage{mathtools}
\usepackage{physics}
\usepackage{amssymb}
\usepackage{tcolorbox}

\begin{document}

\begin{titlepage}

\widetext

\title{Magnetothermopower of nodal line semimetals}

\author{Poulomi Chakraborty}
\affiliation{Department of Physics, The Ohio State University, Columbus, Ohio 43202, USA}
\author{Aaron Hui}
\affiliation{Department of Physics, The Ohio State University, Columbus, Ohio 43202, USA}
\author{Grigory Bednik}
\affiliation{Department of Physics, University of Nebraska, Omaha, Nebraska 68182, USA}
\author{Brian Skinner}
\affiliation{Department of Physics, The Ohio State University, Columbus, Ohio 43202, USA}

\setcounter{equation}{0}
\setcounter{figure}{0}
\setcounter{table}{0}

\makeatletter
\renewcommand{\theequation}{S\arabic{equation}}
\renewcommand{\thefigure}{S\arabic{figure}}
\renewcommand{\thetable}{S\Roman{table}}
\renewcommand{\bibnumfmt}[1]{[S#1]}
\renewcommand{\citenumfont}[1]{S#1}

\date{\today}

\begin{abstract}
The search for materials with large thermopower is of great practical interest. 
Dirac and Weyl semimetals have recently proven to exhibit superior thermoelectric properties, particularly when subjected to a quantizing magnetic field. Here we consider whether a similar enhancement arises in nodal line semimetals, for which the conduction and valence band meet at a line or ring in momentum space. We compute the Seebeck and Nernst coefficients for arbitrary temperature and magnetic field and we find a wealth of different scaling regimes. Most strikingly, when a sufficiently strong magnetic field is applied along the direction of a straight nodal line or in the plane of a nodal ring, the large degeneracy of states leads to a large, linear-in-$B$ thermopower that is temperature-independent even at low temperatures. Our results suggest that nodal line semimetals may offer significant opportunity for efficient, low-temperature thermoelectrics.
\end{abstract}


\maketitle



\end{titlepage} 

\section{Introduction}
\label{sec:Intro}

Charge carriers in a solid material generally also carry heat, so that the phenomena of electrical and thermal conduction become mixed. Under steady-state conditions where no current is flowing, a temperature gradient leads to a gradient of carrier concentration and therefore to a measurable voltage difference; this is known as the thermoelectric effect. 
The strength of the thermoelectric effect is typically characterized by the Seebeck coefficient (thermopower), defined via \cite{Ashcroft, ioffe_semiconductor_1957, shakouri_recent_2011} 
\begin{align}
    S_{xx} = \frac{(\Delta V)_x}{(\Delta T)_x},
    \label{eq: S_xx open circuit def}
\end{align}
where $(\Delta T)_x$ is the difference in temperature along the $x$ direction and $(\Delta V)_x$ is the voltage difference along the same direction. Identifying or designing materials with large Seebeck coefficient has great utility in practical applications such as thermocouples, thermoelectric coolers, and thermoelectric generators, since the thermoelectric effect allows one to convert waste heat to electric current, or to convert applied electric current to heating or cooling power \cite{Snyder2008}.

In typical conductors, however, the thermoelectric effect is generically weak. For example, in a single-band conductor with a large Fermi surface the Seebeck coefficient is of order $(k_B/e) (k_B T/E_F)$ with $E_F$ the Fermi energy and $-e$ the electron charge \cite{Ashcroft, Girvin_Yang_2019,FRITZSCHE19711813}. The quantity $k_B/e \approx 86$\,$\mu$V/K represents the natural unit of thermopower.
Typical metals, for which $k_B T \ll E_F$, therefore have a greatly suppressed thermopower. On the other hand, materials with low $E_F$, such as doped semiconductors or insulators, are poor electrical conductors; they can achieve high thermopower but have high electrical resistance, hampering their ability to provide efficient power conversion \cite{FRITZSCHE19711813,shakouri_recent_2011,Chen2013anomalously}.


The recently-discovered Dirac and Weyl semimetals \cite{armitage_review_2018} are potentially very useful for thermoelectric applications because they offer the promise of arbitrarily low Fermi energy combined with robust electrical conductivity. Particularly notable is a strong magnetothermoelectric effect: a series of recent works 
\cite{peng2016high, wang2018magnetic, Xiang2019, Fu_Felser_review, Skinnereaat2621, KoziiSkinner, scott_doping_2023,Zhang2020, Han2020, Watzman_2018, Liang_2017, Zhu_2015, wangcd3as2} 
has demonstrated that an applied magnetic field can greatly enhance the thermopower of Dirac and Weyl semimetals while still maintaining metallic-like electrical conduction. 


A particularly striking magnetothermoelectric effect was pointed out in Ref.~\cite{Skinnereaat2621}, which showed that a Dirac or Weyl semimetal can exhibit a nonsaturating linear-in-$B$ growth of the Seebeck coefficient when the magnetic field $B$ is large enough to achieve the extreme quantum limit (EQL), in which all electrons reside in the lowest Landau level. 
The crux of this effect lies in the way that a large magnetic field increases the electronic entropy. In the limit where the magnetic field is large enough to produce a large Hall angle (i.e., where electric current flows nearly perpendicular to the electric field), the Seebeck coefficient becomes directly proportional to the electronic entropy \cite{obraztsov1965thermal,qmtheorythermophenomena,tsendin1966theory,jay-gerin_thermoelectric_1974,Girvin_1982,bergman_dissipationless_nernst,abrikosov2017fundamentals,Skinnereaat2621, KoziiSkinner}:
\begin{align}
    S_{xx} = \frac{n_s}{\rho_e},
    \label{eq: Seebeck entropy relation}
\end{align}
where $n_s$ is the entropy density and $\rho_e$ is the charge density of mobile carriers. At low temperatures the entropy density of an electronic system is directly proportional to the density of states $\nu$ near the Fermi level: $n_s \simeq (\pi^2/3) k_B^2 T \nu(E_F)$. Since the degeneracy of a given Landau level, and therefore the value of $\nu$, grows linearly as a function of $B$, the Seebeck coefficient $S_{xx}$ grows linearly in $B$ when only a single Landau level is occupied. This linear-in-$B$ enhancement of the Seebeck coefficient was observed experimentally in Refs.\ \onlinecite{liang2013evidence, Zhang2020, Han2020, scott_doping_2023}, for example. Notice, however, that $S_{xx}$ vanishes as $T \rightarrow 0$, such that achieving a large thermopower at low temperature requires a very large magnetic field.

\begin{figure}
    \centering
    \includegraphics[width=\columnwidth]{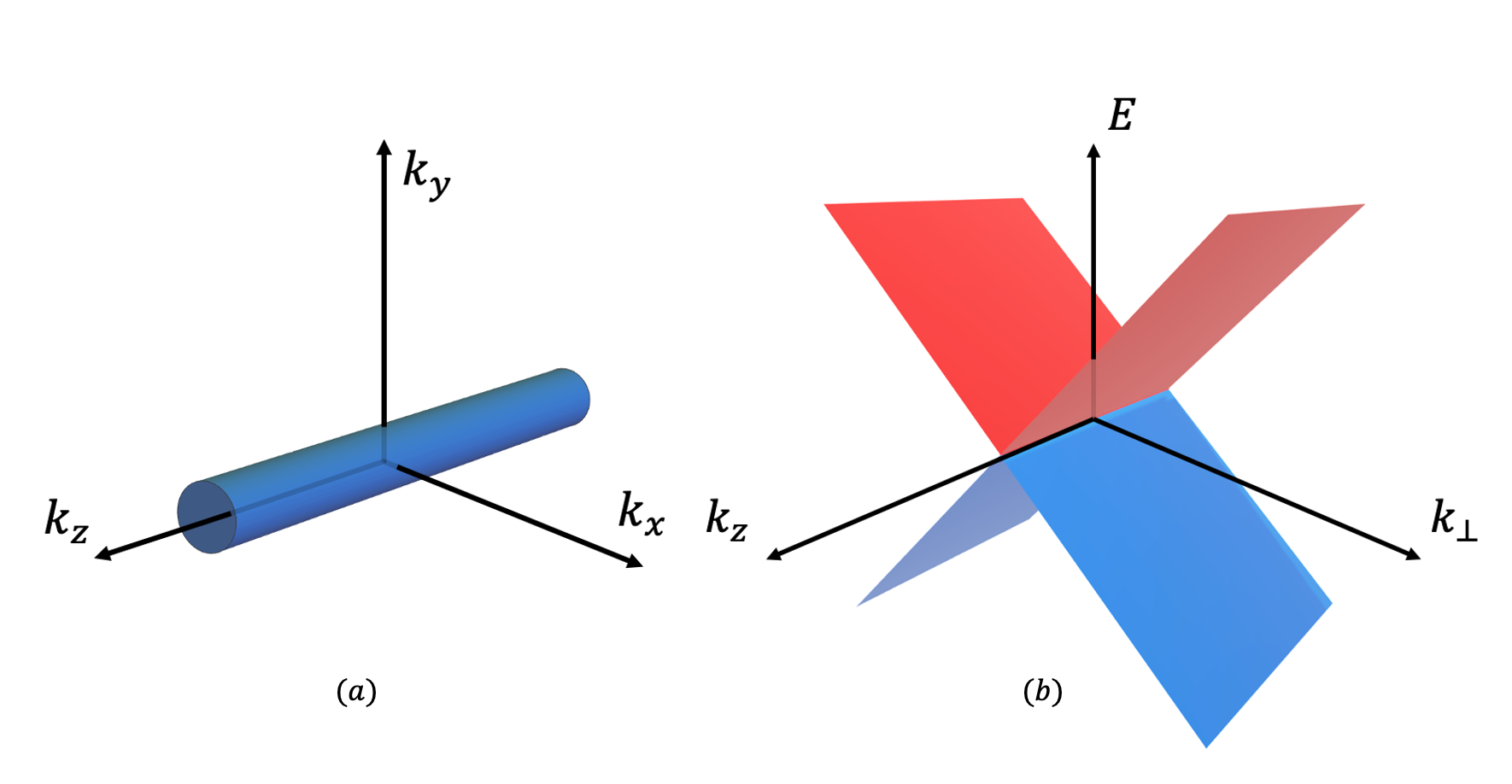}
    \caption{(a) Schematic of a cylindrical Fermi surface. (b) The plot of the dispersion relation in the $k_x - k_y$ plane, i.e. perpendicular to the direction of the nodal line.  }
    \label{fig: straightnodalline}
\end{figure}

In this paper we consider whether a similar magnetic-field enhancement of thermopower can appear in nodal line semimetals (NLSs), for which gapless points in the dispersion relation form a 1D manifold such as a line (illustrated in Fig.\ \ref{fig: straightnodalline}) or a ring in momentum space \cite{NLSM_Fang_2016, NLSM_Burkov, NLSM_Zhong, NLSM_Chen2015-lu, nodal_line_transport}. We focus on the case where the nodal line exists at a constant energy, and we calculate the longitudinal and transverse thermopower (the Seebeck and Nernst coefficients) as a function of both temperature and magnetic field across the entire range of both parameters. We find that the thermopower generally exhibits a strong enhancement with magnetic field. Most prominently, the thermopower rises linearly with $B$ in the extreme quantum limit, as for Dirac and Weyl semimetals, but for NLSs the flat dispersion along the nodal line enables a huge electronic entropy even at low temperature. Consequently, the Seebeck coefficient in the extreme quantum limit is large, linear-in-$B$, and temperature-independent at low temperature. This effect may lead to practical low-temperature thermoelectrics. For example, as we show below, in the realistic case of a NLS with electron density $n = 10^{18}$\, cm$^{-3}$, Fermi velocity $v_F = 10^5 $\,m/s, and a circular nodal line with radius $0.1$ \AA$^{-1}$, increasing the strength of an applied magnetic field from $B = 0$ to $B = 10$\,T can increase the thermopower at $T = 10$\,K from $S_{xx} \approx 40$\,$\mu$V/K to $S_{xx} \approx 300$\,$\mu$V/K. At lower doping, 
the effect is even more dramatic: a NLS with $n = 10^{17}$\,cm$^{-3}$ sees its thermopower increase from $S_{xx} \approx 150$\,$\mu$V/K to $S_{xx} \approx 3000$\,$\mu$V/K under the same conditions.

The remainder of this paper is organized as follows. In Sec.~\ref{sec: approaches} we explain our mathematical setup and we show how to calculate the Seebeck and Nernst coefficients at arbitrary $B$ and $T$ using two complementary approaches. In Sec.~\ref{sec: straight}, we discuss the results for a straight nodal line, considering each regime of $B$ and $T$. In Sec.~\ref{sec: circular}, we discuss the generalization of our results to the case of a circular nodal line. We conclude in Sec.~\ref{sec: conclusion} with a summary and a brief discussion of NLS materials and limitations of our theory.

\section{Calculational Approach}
\label{sec: approaches}

\subsection{Setup and Definitions}

Our goal in this paper is to compute the thermoelectric tensor $\hat{S}$ as a function of magnetic field $B$ and temperature $T$; the diagonal component of $\hat{S}$ is the Seebeck coefficient (in a particular direction) and the off-diagonal component is the Nernst cofficient. The tensor $\hat{S}$ can be defined by considering the equations that govern the electrical and heat current densities, $\mathbf{J}^E$ and $\mathbf{J}^Q$:
\begin{align}
    \mathbf{J}^E &= \hat{\sigma}\mathbf{E} - \hat{\alpha}\grad{T}
    \label{eq: je}
    \\
    \mathbf{J}^Q &= T\hat{\alpha}\mathbf{E} - \hat{\kappa}\grad{T}.
    \label{eq: jq}
\end{align}
Here, $\mathbf{E}$ is the electric field and $\hat{\sigma}, \hat{\alpha}, \hat{\kappa}$ are the electrical, Peltier, and thermal conductivity tensors respectively. The appearance of the same coefficient $\hat{\alpha}$ in the ``off-diagonal'' term of both Eqs.~\eqref{eq: je} and \eqref{eq: jq} is a reflection of Onsager reciprocity \cite{Ashcroft}. The thermoelectric tensor is defined by $\mathbf{E} = \hat{S} \grad{T}$ under conditions where the electrical current $\mathbf{J}^E = 0$ [as in Eq.~\eqref{eq: S_xx open circuit def}], and therefore
\begin{align}
    \hat{S} = \hat{\sigma}^{-1}\hat{\alpha}.
    \label{generalS}
\end{align}
Writing out the Seebeck $S_{xx}$ and Nernst $S_{xy}$ coefficients explicitly gives
\begin{align}
    S_{xx} & = \frac{ \alpha_{xx}\sigma_{xx} + \alpha_{xy}\sigma_{xy}}{\sigma_{xx}^2 +\sigma_{xy}^2}
    \label{eq: sxx}
    \\
    S_{xy} &= \frac{ \alpha_{xy}\sigma_{xx} - \alpha_{xx}\sigma_{xy}}{\sigma_{xx}^2 +\sigma_{xy}^2}.
    \label{eq: sxy}
\end{align}

Following Refs.~\cite{KoziiSkinner, xiaozhoufeng}, we calculate $\hat{S}$ across the full range of $B$ and $T$ by using two complementary approaches. When many Landau levels are occupied (i.e., when the Landau level spacing is small compared to the Fermi energy), we can use a semiclassical approach to directly calculate the electrical and Peltier conductivity tensors $\hat{\sigma}$ and $\hat{\alpha}$. In the extreme quantum limit, however, where only a single Landau level is occupied, direct calculation of these tensors becomes difficult. The electron dispersion relation is strongly modified by Landau quantization and the mobility in the field direction becomes strongly different from the mobility in the perpendicular directions. Calculating $\hat{\alpha}$ is particularly subtle, requiring one to invoke Landau level broadening effects.

Fortunately, at sufficiently large magnetic field one can use the generic expression given by Eq.~\eqref{eq: Seebeck entropy relation}, which only requires one to calculate the thermodynamic entropy of the electron system. This formula is valid whenever $\sigma_{xy} \gg \sigma_{xx}$; we refer to this limit as the ``dissipationless" limit 
since the dissipative conductivity $\sigma_{xx}$ is negligible. Our two calculational schemes -- the semiclassical and dissipationless limits -- have an overlapping regime of validity, and we show below that both approaches agree quantitatively in the overlap regime.

Throughout this paper we assume that the charge density of carriers in the nodal band is constant. That is, we take $n_e - n_h = \textrm{constant}$, where $n_e$ and $n_h$ are the concentration of electron- and hole-type carriers, respectively, in the nodal bands. Both $n_e$ and $n_h$ can in general depend on $T$ and $B$. This assumption of constant charge density is natural when there are no other trivial bands nearby in energy. In cases where another band with large density of states coexists in energy, one can instead have a condition where the chemical potential $\mu$ in the nodal band is constant as a function of $T$ and $B$ due to the large trivial band acting as a reservoir. Such pinning of the chemical potential affects the temperature dependence of the thermopower when $T$ is much larger than the Fermi temperature $T_F$, leading to quantitative but not qualitative changes to the results we discuss below.


\subsection{Semiclassical limit}












In the limit of sufficiently small magnetic field 
that many Landau levels are occupied, we can ignore Landau quantization effects and calculate the electrical and Peltier conductivity tensors via the usual semiclassical approach \cite{Ashcroft}:
\begin{align}
    \sigma_{ij} &= \int d\epsilon \left( -\frac{\partial f}{\partial \epsilon} \right) \sigma_{ij} (\epsilon) 
    \label{eq: sigma conductivity}
    \\
    \alpha_{ij} &= \frac{1}{eT} \int d\epsilon  (\epsilon - \mu) \left( -\frac{\partial f}{\partial \epsilon} \right) \sigma_{ij} (\epsilon) 
    \label{eq: alpha conductivity}.
\end{align}
Here, $\sigma_{ij}(\epsilon)$ corresponds to the conductivity of carriers at energy $\epsilon$ (i.e., to the zero-temperature conductivity when the chemical potential $\mu$ is equal to $\epsilon$) and $f(\epsilon) \equiv \left( \exp\left(\frac{\epsilon - \mu}{k_B T}\right) +1 \right)^{-1}$ is the Fermi-Dirac distribution. Thus, Eq.~\eqref{eq: sigma conductivity} corresponds to a weighted average of quasiparticle contributions to conductivity in a thermally-broadened window around the chemical potential. Similarly, $\alpha_{ij}$ corresponds to a weighted-average of quasiparticle contributions to the thermal current under an applied electric field, and therefore includes both $(\epsilon - \mu)$ energy and $\sigma_{ij}$ electrical factors. At low temperatures $T \ll T_F$, a Sommerfeld expansion directly yields the usual Mott formula \cite{Ashcroft}
\begin{align}
    \hat{S} = \frac{k_B}{e}\frac{\pi^2}{3} k_B T \hat{\sigma}(\epsilon)^{-1} \left.\frac{d\hat{\sigma}(\epsilon)}{d\epsilon} \right|_{\epsilon = \mu}.
    \label{eq: Mott formula}
\end{align}
We remark that the Mott formula is generic at low temperature $T \ll T_F$, independent of particular assumptions about $\sigma_{ij}(\epsilon)$. It breaks down only when either $T \gg T_F$, such that one can no longer use a Sommerfeld expansion for the integrals in Eqs.~\eqref{eq: sigma conductivity} and (\ref{eq: alpha conductivity}), or when $\hat{\sigma}$ becomes discontinuous as a function of chemical potential (such as in the regime of strong Landau quantization).


Throughout the main text of this paper we assume that the chemical potential $\mu$ is determined self-consistently from the condition of constant charge density within the nodal line-bands, i.e.,
\begin{align}
    n_e - n_h = \int_0^{\infty} d\epsilon \, \nu (\epsilon) f(\epsilon) - \int_{-\infty}^0 d\epsilon \, \nu (\epsilon)(1- f(\epsilon)) = \textrm{const.}
    \label{eq: mu vs T}
\end{align}
The first term on the right-hand side of this expression represents the number of electrons in the conduction band, while the second term describes the number of holes in the valence band.
At low temperatures $(T \ll T_F)$, the chemical potential $\mu \simeq E_F$. At high temperatures $(T \gg T_F)$, evaluating Eq.~\eqref{eq: mu vs T} gives a chemical potential $\mu \simeq E_F^2/(4 k_BT \ln{2})$, assuming low enough temperature that the typical quasiparticle momentum is small compared to the diameter of the nodal line.  We discuss in Appendix \ref{app: constantmu} how our results are modified if $\mu$ is taken to be constant as a function of temperature, as may result from pinning by a trivial electron band that coexists in energy with the nodal-line bands.

In Sec.~\ref{sec: straight} we give special consideration to the case of a straight nodal line, for which the bands are dispersionless along the $k_z$ direction. In this case only the component $\mathbf{B} \cdot \mathbf{\hat{z}} \equiv B$ of the magnetic field along the $z$ direction is relevant, since electrons have infinite mass along the $z$ direction. 
In order to calculate the tensors $\hat{\sigma}$ and $\hat{\alpha}$ in this case, we use the relaxation-time approximation with a momentum relaxation time $\tau$. In this case the energy-dependent conductivities $\sigma_{ij} (\epsilon)$ are given by
\begin{align}
    \sigma_{xx} (\epsilon) &= \frac{\nu_\text{2D}(\epsilon)}{2} \frac{e^2 v_F^2 \tau}{1+\omega_c^2(\epsilon)\tau^2}
    \label{eq: sigma xx}
    \\
    \sigma_{xy} (\epsilon) &= \frac{\nu_\text{2D}(\epsilon)}{2} \frac{e^2 v_F^2 \tau}{1+\omega_c^2(\epsilon)\tau^2} \omega_c(\epsilon) \tau .
    \label{eq: sigma xy}
\end{align}
Here, 
\be 
\omega_c (\epsilon) = \frac{eBv_F^2}{\epsilon}
\label{eq: omegac}
\ee 
is the energy-dependent cyclotron frequency, with $v_F$ the Fermi velocity in the direction perpendicular to the magnetic field and 
\be 
\nu_\text{2D}(\epsilon) = \frac{g}{2\pi} \frac{|\epsilon|}{\hbar^2 v_F^2}
\label{eq: DoS}
\ee 
being the density of states per value of the momentum along the nodal line. The constant $g$ denotes the degeneracy per momentum state (spin $\times$ valley). The full, three-dimensional density of states $\nu(\epsilon)$ is equal to $\nu_\text{2D}(\epsilon)$ multiplied by the length of the nodal line in reciprocal space. Notice that $\omega_c$ has the same sign as the energy $\epsilon$, so that electrons and holes have opposite signs of the cyclotron frequency.
For simplicity, we assume throughout this paper that $\tau$ is independent of $\epsilon$; incorporating a power-law energy dependence into $\tau$ changes various order-1 numerical coefficients but does not affect the way that $\hat{S}$ scales with $T$, $B$, $\tau$, or $E_F$ \cite{hwang_theory_2009}.



\subsection{Dissipationless limit}
\label{subsec: Dissipationless limit}







In the dissipationless limit $\sigma_{xy} \gg \sigma_{xx}$, the Seebeck coefficient is given directly by Eq.~\eqref{eq: Seebeck entropy relation}, such that for a given charge density one need only calculate the thermodynamic entropy. The entropy per unit volume is generically given by \cite{Landau_Lifshitz_1980,bergman_dissipationless_nernst,KoziiSkinner} 
\begin{align}
    n_s =& -k_B \sum_n \int d \epsilon_n \; \nu(\epsilon_n) \Big[ f(\epsilon_n) \ln f(\epsilon_n)
    \nonumber
    \\
    &\phantom{\sum_n \int d \epsilon_n \; \nu(\epsilon_n)}  + (1 - f (\epsilon_n)) \ln (1 - f(\epsilon_n)) \Big],
    \label{eq: entropy density}
\end{align} 
where the sum is taken over the Landau level index $n$. \cite{Girvin_1982, Halperin, KoziiSkinner} One can view Eq.~\eqref{eq: entropy density} as the Shannon entropy $-k_B [f \ln f + (1-f) \ln (1-f)]$ for a quantum state that is occupied with probability $f$, integrated over all quantum states.
Thus, calculating the Seebeck coefficient in this high-$B$ limit requires only knowledge of the density of states $\nu(\epsilon)$ and the Landau level spectrum $\epsilon_n$. In this paper we take $\nu(\epsilon)$ to be that of the noninteracting band structure, essentially ignoring electron-electron interactions in our calculation of the entropy. This assumption fails  at sufficiently low temperatures, as we discuss in Sec.~\ref{sec: conclusion}.

For the straight nodal line with magnetic field $B$ applied along the axis of the nodal line, which is our primary consideration in this paper, the expressions for $\nu$ and $\epsilon_n$ are \cite{Geim2007, KoziiSkinner}
\begin{align}
    \epsilon_n (B) &= \textrm{sign}(n) v_F \sqrt{2e\hbar B \abs{n}}
\\
    \nu(\epsilon) &= \frac{eB}{\hbar} k_0 g \sum_n \delta (\epsilon - \epsilon_n),
\end{align}
where $2 \pi k_0$ is the $k$-space length of the nodal line. (We denote the length of the nodal line as $2 \pi k_0$ for reasons of notational consistency with the case of a circular nodal line considered in Sec.~\ref{sec: circular}.)

\section{Straight nodal line}
\label{sec: straight}

\begin{figure*}[htb]
    \centering
    \includegraphics[width=.8\textwidth]{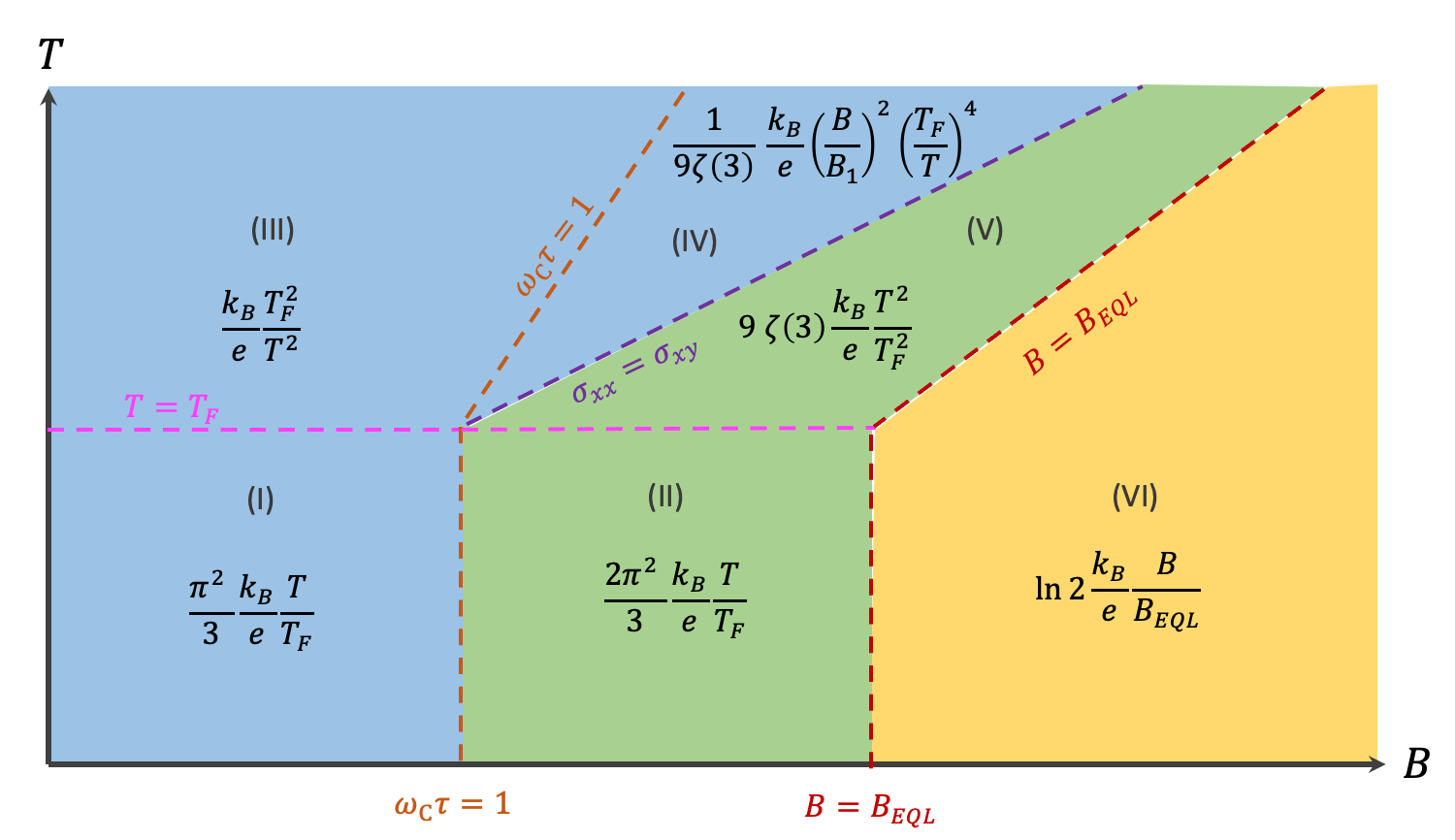}
    \caption{A schematic plot of the Seebeck coefficient $S_{xx}$ versus $B$ and $T$ for a NLS with a straight nodal line in various asymptotic regimes. The two axes are depicted on a logarithmic scale. Blue-shaded regions correspond to regimes where $S_{xx}$ is calculated using a semiclassical approach, yellow regions correspond to regimes for which $S_{xx}$ is calculated using the dissipationless limit approach, and green regions correspond to regimes where both approaches are valid and give the same result. The various dashed lines denote the relevant temperature and magnetic field scales that define the different regimes. $B_1$ and $B_{\textrm{EQL}}$ are defined in Eq.(\ref{eq: B1 definition}) and (\ref{eq: B_EQL definition}).}
    \label{fig: sxxmap}
\end{figure*}

In this section we focus on the case of a NLS with a straight nodal line. That is, we consider an electronic system with the Hamiltonian
\begin{align}
    H = v_F \left(\sigma_x p_x + \sigma_y p_y + \sigma_z F(p_z) \right),
    \label{eq: Hstraight}
\end{align}
where $v_F$ is the Fermi velocity (Dirac velocity), which we take to be a constant, $\sigma_i$ are the Pauli matrices, and $p_i$ is the momentum in direction $i$. The nodal line resides along the $p_z$ direction, and the function $F(p_z)$ produces a variation in energy (``corrugation'') of the nodal line. For the remainder of this paper we take $F(p_z) \equiv 0$, so that the nodal line resides at a single energy. We discuss the effects of corrugation of the nodal line in Sec.\ \ref{sec: conclusion}.  A schematic Fermi surface and dispersion relation are plotted in Fig.~\ref{fig: straightnodalline}. The Hamiltonian \eqref{eq: Hstraight} with $F(p_z) \equiv 0$ is equivalent to that of graphene, and therefore one can think of a NLS with a straight and flat nodal line as equivalent to a stack of noninteracting 2D graphene layers. In this case, as mentioned above, only the magnetic field component in the stacking direction ($z$ direction) has any influence on the electronic system (neglecting Zeeman coupling to the electron spin), and therefore we assume without loss of generality that the magnetic field $\mathbf{B} = B\hat{\mathbf{z}}$ is aligned along the nodal line.

Before presenting results for the thermopower, we pause to discuss the physical scales associated with the problem; these are demarcated by the various regions in Fig.~\ref{fig: sxxmap}. At finite electron concentration $n_e$, a NLS at zero temperature has a Fermi energy $E_F = \hbar v_F k_F$, where $k_F = \sqrt{4 \pi n_e/(g k_0)}$ is the Fermi wave vector. The Fermi energy sets the characteristic temperature $T_F = E_F/k_B$ of the electronic system. Thus we can first partition our parameter space by comparing $T$ against $T_F$.

At low temperature $T \ll T_F$, there are two characteristic magnetic field scales: a scale $B_1$ at which $\omega_c(\epsilon = E_F) \tau = 1$ and a scale $B_\text{EQL}$ above which only the lowest Landau level is occupied. Using Eq.~\eqref{eq: omegac} gives 
\begin{align}
    B_1 = \frac{E_F}{e v_F^2 \tau}.
    \label{eq: B1 definition}
\end{align}
At fields $B \gg B_1$ one can think that the conductivity tensor is strongly modified from its $B = 0$ value and $\sigma_{xy} \gg \sigma_{xx}$. Calculationally, in the limit $B \gg B_1$ we can simplify the denominator of Eq.~\eqref{eq: sigma xx} and Eq.~\eqref{eq: sigma xy} when performing asymptotic estimates of $S_{ij}$. The larger field scale 
\begin{align}
B_\text{EQL} = \frac{2 \pi \hbar n_e}{e g k_0} = \frac{E_F^2}{2 e \hbar v_F^2} = B_1 \frac{E_F \tau}{\hbar}
\label{eq: B_EQL definition}
\end{align}
can be identified by the condition that the degeneracy of the lowest Landau level exceeds the electron concentration. The two field scales are well separated, $B_\text{EQL} \gg B_1$, so long as the scattering rate $1/\tau \ll E_F/\hbar$, which is a generic requirement for having a conductor with well-defined quasiparticles. Thus at $T \ll T_F$ there are three regimes of magnetic field: a low field regime $B \ll B_1$, an intermediate field regime $B_1 \ll B \ll B_\textrm{EQL}$, and a high field regime $B \gg B_{EQL}$.

At high temperatures $T \gg T_F$, both field scales $B_1$ and $B_\text{EQL}$ are modified due to the energy-dependence of $\omega_c$ and the proliferation of thermally-excited holes in the valence band. Notably, the large concentration of holes implies that the Hall conductivity $\sigma_{xy}$ is significantly reduced for a given field strength relative to its $T=0$ value, since the contribution of holes to $\sigma_{xy}$ has the opposite sign as the electron contribution. Thus, at $T \gg T_F$ the field scale $B_1$ splits into two different values: a smaller value associated with $\omega_c(\epsilon = k_B T) \tau = 1$, since the quasiparticles have characteristic energy $\epsilon \sim k_B T$, and a larger value associated with $\sigma_{xx} = \sigma_{xy}$ [where $\sigma_{xx}$ and $\sigma_{xy}$ refer to the thermally-averaged values of the conductivity, given by Eq.~\eqref{eq: sigma conductivity}, and not to the values at a specific fixed energy, given by Eqs.~\eqref{eq: sigma xx} and \eqref{eq: sigma xy}].  The smaller value associated with $\omega_c(\epsilon = k_B T) \tau = 1$ is given by $B \sim (T/T_F) B_1$, while the larger value associated with $\sigma_{xx} = \sigma_{xy}$ is given by $B \sim (T/T_F)^3 B_1$. The onset of the EQL is also delayed to $B \sim (T/T_F)^2 B_\text{EQL}$ at such high temperatures, since it corresponds to the condition where the Landau  level spacing is larger than $k_B T_F$ rather than $E_F$. Thus at $T \gg T_F$ there are four distinct regimes of magnetic field.




The different regimes of field and temperature are summarized in Fig.~\ref{fig: sxxmap}, along with the corresponding asymptotic relation for the Seebeck coefficient $S_{xx}$. In the remainder of this section we summarize the behavior in each of these regimes.

\subsection{Seebeck Coefficient}

We now compute the Seebeck coefficient $S_{xx}$ across the various regimes outlined above (see Fig.~\ref{fig: sxxmap}). In the limit of low temperature $T \ll T_F$ and outside the EQL ($B \ll B_\text{EQL}$), which corresponds to regions (I) and (II) in Fig.~\ref{fig: sxxmap}, one can calculate the Seebeck coefficient using the usual Mott formula [Eq.~\eqref{eq: Mott formula}]. These calculations give
\be 
   S_{xx}^\text{I} \simeq \frac{\pi^2}{3} \frac{k_B}{e} \frac{T}{T_F}
   \label{eq: SxxI}
\ee 
for the regime of $\omega_c(\epsilon = E_F) \tau \ll 1$ (small Hall angle), and
\be 
   S_{xx}^\text{II} \simeq \frac{2\pi^2}{3} \frac{k_B}{e} \frac{T}{T_F}
   \label{eq: SxxII}
\ee 
for the regime of $\omega_c(\epsilon = E_F) \tau \gg 1$ (large Hall angle). The exact numerical prefactors in these expressions are dependent on our assumption of an energy-independent scattering time $\tau$; introducing a power-law dependence $\tau(\epsilon)$ would introduce different numeric prefactors to Eqs.~\eqref{eq: SxxI} and \eqref{eq: SxxII} \cite{hwang_theory_2009}. But the proportionality of the Seebeck coefficient to $T/T_F$ is universal. One can think that this dependence arises because the electronic entropy is vanishingly small at low temperature: at low temperature only a small fraction $\sim T/T_F$ of electrons contribute significantly to the entropy, so that the ratio of entropy to charge is also proportional to $T/T_F$.

\begin{figure}
    \centering
    \includegraphics[width=\columnwidth]{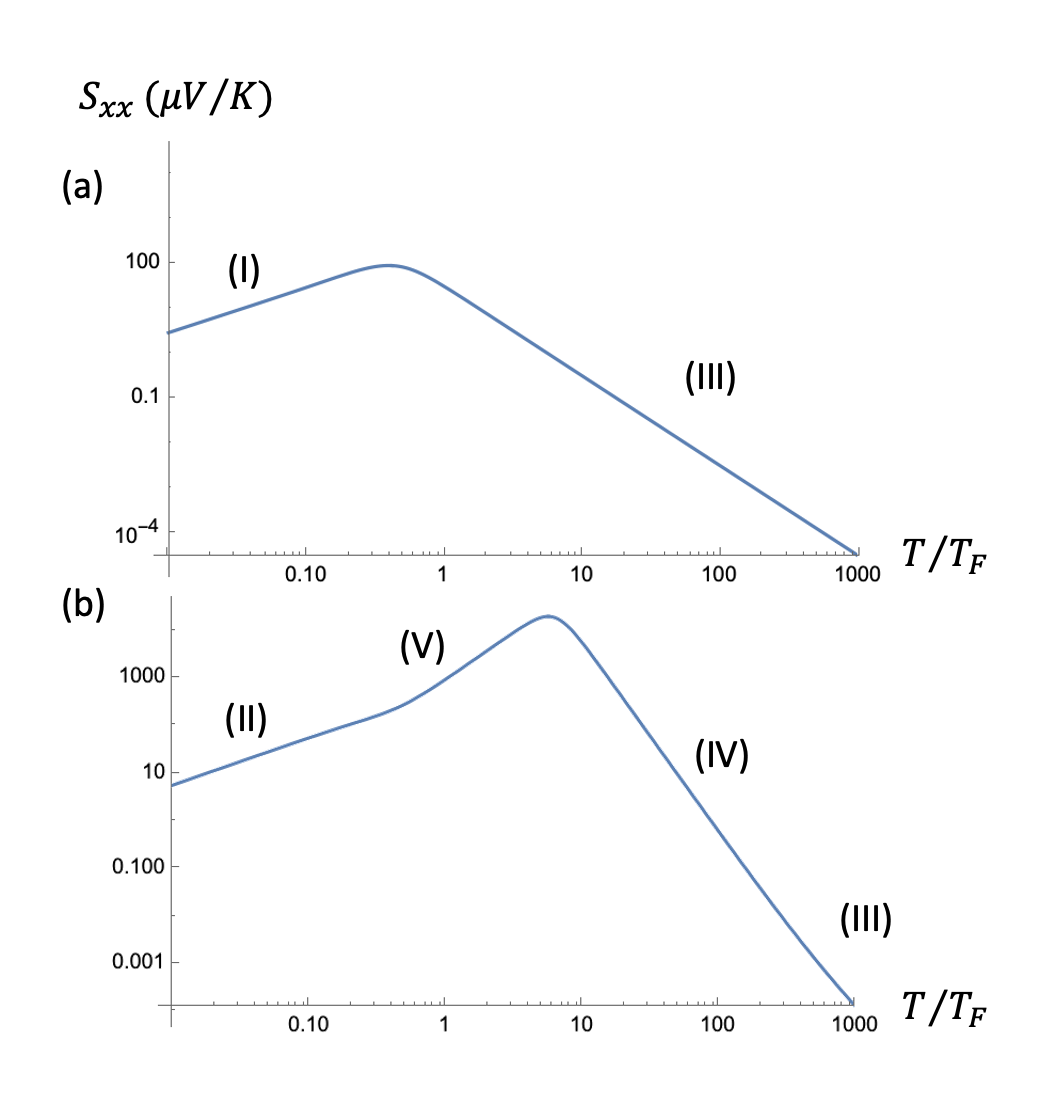}
    \caption{(a) Variation of the Seebeck coefficient with temperature at a weak magnetic field $B = 0.1 B_1$ (i.e., $\omega_c \tau \ll 1$). 
     (b) Variation of the Seebeck coefficient with temperature at a strong magnetic field $B = 27 B_1$ (i.e., $\omega_c \tau \gg 1$ at $T \ll T_F$). Results are calculated using Eqs.~\eqref{eq: sxx}, \eqref{eq: sigma conductivity}, and \eqref{eq: alpha conductivity}.
     }
    \label{fig: tlinecuts}
\end{figure}

At high temperatures $T \gg T_F$ there are many thermally excited electron-hole pairs, so that the total carrier density $n_e + n_h \sim k_0 (k_B T)^2/(\hbar v_F)$ greatly exceeds its value at low temperature. In this way the longitudinal conductivity $\sigma_{xx}$ is greatly enhanced by temperature. The longitudinal Peltier conductivity $\alpha_{xx}$, on the other hand, is greatly reduced by increased temperature. In the absence of magnetic field, electrons and holes move in opposite direction under the presence of an electric field, and consequently they carry heat in opposite directions so that their contributions to $\alpha_{xx}$ cancel. Evaluating Eqs.~\eqref{eq: sxx} and \eqref{eq: alpha conductivity} in the limit of high temperature and low magnetic field (region III in Fig.~\ref{fig: sxxmap}), one arrives at
\begin{align}
   S_{xx}^\text{III} \simeq \frac{k_B}{e} \frac{T_F^2}{T^2}.
   \label{eq: SxxIII}
\end{align}
Comparing Eqs.~\eqref{eq: SxxI} and \eqref{eq: SxxIII}, one can see that at low magnetic field the Seebeck coefficient first increases then decreases with temperature, achieving a maximum of order $k_B/e$ at a temperature $T \sim T_F$. This behavior is shown in Fig.~\ref{fig: tlinecuts}(a), which gives a full calculation of $S_{xx}(T)$ at low $B$.

\begin{figure}
    \centering
    \includegraphics[width=\columnwidth]{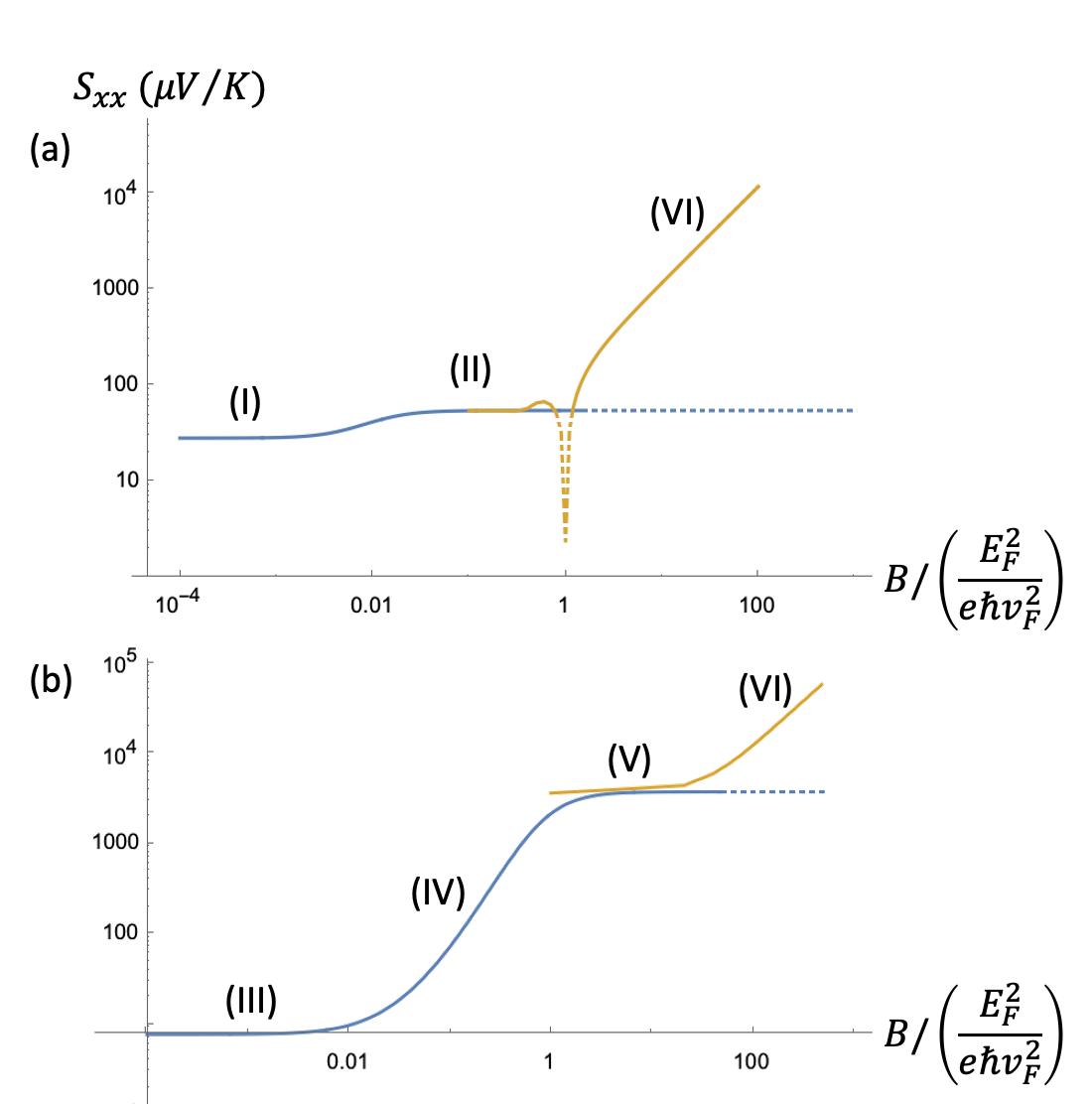}
    \caption{(a) Variation of the Seebeck coefficient with magnetic field at a low temperature $T = 0.1 T_F$. The unit of magnetic field on the $x$ axis is equal to $2B_\text{EQL}$.
    (b) Variation of the Seebeck coefficient with magnetic field at a high temperature $T = 2 T_F$. Blue lines in this plot (regimes I, II, III, IV, and V) are calculated using the semiclassical approach of Eqs.~\eqref{eq: sxx}, \eqref{eq: sigma conductivity}, and \eqref{eq: alpha conductivity}, while the yellow lines (regimes II, V and VI) correspond to the dissipationless limit calculation [Eqs.~\eqref{eq: Seebeck entropy relation} and \eqref{eq: entropy density}]. In both plots the value of the momentum relaxation time is taken to be $\tau = 100 \hbar/E_F$.}
    \label{fig: blinecuts}
\end{figure}

As one turns on $B$ at high temperatures $T \gg T_F$ (see Fig.~\ref{fig: blinecuts}), the electrical conductivity tensor is modified due to the Lorentz force, and at sufficiently large $B$ that $\omega_c(\epsilon = k_B T) \tau \gg 1$ it becomes strongly modified. Within this large-field limit, the distinct regimes IV and V in Fig.~\ref{fig: sxxmap} are demarcated by whether $\sigma_{xx}$ remains much larger than $\sigma_{xy}$. 
In regime IV we find
\begin{align}
    S_{xx}^\text{IV} \simeq \frac{1}{9\zeta(3)} \frac{k_B}{e} \left(\frac{B}{B_1} \right)^2 \left(\frac{T_F}{T}\right)^4,
    \label{eq: SxxIV}
\end{align}
where $\zeta(x)$ is the Riemann zeta function.
Equation \eqref{eq: SxxIV} implies a strong, $B^2$ enhancement of the thermopower by magnetic field. Conceptually, this strong enhancement arises because electron and hole type carriers are able to carry heat in parallel via a longitudinal component of the $\mathbf{E} \times \mathbf{B}$ drift \cite{xiaozhoufeng}.

In regime V, where $\sigma_{xy} \gg \sigma_{xx}$ (the ``dissipationless limit"), we find 
\begin{align}
    S_{xx}^\text{V} \simeq 9 \zeta (3) \frac{k_B}{e} \frac{T^2}{T_F^2}.
    \label{eq: SxxV}
\end{align}
Notice that at such large fields the Seebeck coefficient reaches a plateau value that is enormously enhanced relative to the $B = 0$ value by a factor $(T/T_F)^4$. Such a large enhancement arises due to the large Hall effect, which allows electron and hole carriers to contribute additively to the thermopower, since their motion is governed by the $\mathbf{E} \times \mathbf{B}$ drift, rather than canceling as they do at small field \cite{Skinnereaat2621, xiaozhoufeng}. Within regime V the thermopower increases quadratically with temperature, arising from the strongly increasing electronic entropy. Indeed, at such large temperatures the electron system resembles a two-component plasma with nearly equal concentrations of electrons and holes. Since the thermally populated density of carriers $n_e + n_h$ increases as $T^2$, the Seebeck coefficient does as well. The strong enhancement of $S_{xx}$ by magnetic field is shown in Figs.~\ref{fig: tlinecuts}(b) and \ref{fig: blinecuts}(b).

For sufficiently large $B \gg B_\text{EQL}$, only one (dispersionless) Landau level is occupied. At $B \gg B_{\text{EQL}}$ this lowest Landau level approaches half-filling, since it is electron-hole degenerate and the hole states are filled. Thus, $S_{xx}$ is completely determined by the configurational entropy associated with half-filling a highly degenerate Landau level [e.g., Eq.~\eqref{eq: Seebeck entropy relation}]. Using the Landau level degeneracy $N_{\textrm{flat}} = gk_0/l_B^2$ per volume, we can write the entropy density 
\begin{align}
    n_s = k_B \ln{{ N_{\textrm{flat}} \choose \frac{N_{\textrm{flat}}}{2}}} \simeq k_B N_{\textrm{flat}} \ln{2} = k_B \frac{g k_0}{l_B^2} \ln{2}.
\end{align}
Here $l_B = \sqrt{\hbar/(eB)}$ is the magnetic length.
Since the entropy scales linearly with $B$, it is clear that the Seebeck coefficient scales as $S_{xx} \propto B$ with no $T$-dependence. More explicitly, we have
\begin{align}
    S_{xx}^\text{VI} \simeq  2 \ln{2} \frac{k_B}{e} \; \frac{e v_F^2 \hbar}{E_F^2} B = \ln{2} \frac{k_B}{e} \frac{B}{B_\text{EQL}}.
    \label{eq: SxxVI}
\end{align}
The strong enhancement of $S_{xx}$ upon entering the EQL is shown in Fig.~\ref{fig: blinecuts}. The brief, sharp dip in $S_{xx}(B)$ that can be seen in Fig.~\ref{fig: blinecuts}(a) at $B \sim B_{EQL}$ arises because of a suppression of the electronic entropy when the chemical potential lies in the gap between the lowest and second-lowest Landau levels. We note that in deriving Eq.~\eqref{eq: SxxVI} we have implicitly assumed that the density of states is well-described by a noninteracting electron picture (as mentioned in Sec.~\ref{subsec: Dissipationless limit}). At sufficiently low temperatures this assumption becomes invalidated due to exchange splitting of the lowest Landau level. We discuss this issue in Sec.~\ref{sec: conclusion}.

\subsection{Nernst Coefficient}

\begin{figure*}
    \centering
    \includegraphics[width=0.7 \textwidth]{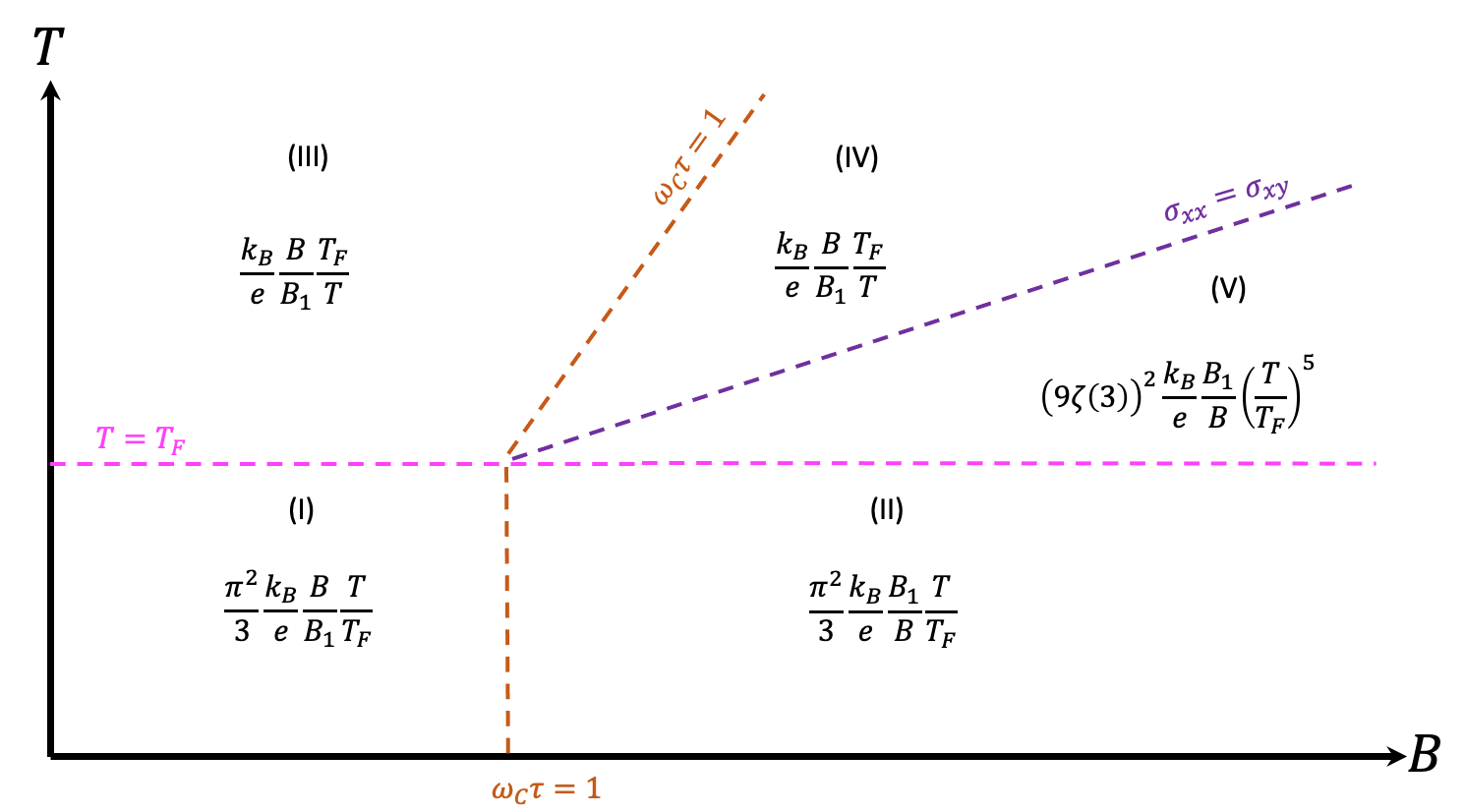}
    \caption{A schematic diagram showing asymptotic expressions for the Nernst coefficient $S_{xy}$ within different regimes of temperature and magnetic field. The labels for each regime are the same as in Fig.~\ref{fig: sxxmap}.  The magnetic field scale $B_1$ is defined by Eq.~(\ref{eq: B1 definition}).}
    \label{fig: sxy map}
\end{figure*}

Let us now summarize the behavior of the Nernst coefficient $S_{xy}$ as a function of $T$ and $B$. We limit our discussion here to the semiclassical regime, where $S_{xy}$ can be computed using Eqs.~\eqref{eq: sxy}, \eqref{eq: sigma conductivity}, and \eqref{eq: alpha conductivity}. Unlike the Seebeck coefficient, the Nernst coefficient is not well defined in the dissipationless limit $\tau \rightarrow \infty$, since the value of $S_{xy}$ always depends on a longitudinal electrical or thermoelectric conductivity [$\sigma_{xx}$ or $\alpha_{xx}$, see Eq.~\eqref{eq: sxy}]. Such longitudinal conductivities cannot be defined without reference to a momentum relaxation mechanism (unlike Hall conductivities, which remain finite even when there is no scattering due to the $\mathbf{E} \times \mathbf{B}$ drift of carriers). However, we show here that the peak value of the Nernst coefficient arises at fields corresponding to $\sigma_{xy} \sim \sigma_{xx}$, which are well below $B_\text{EQL}$. Thus, increasing $B$ into the EQL does not provide additional benefit for the Nernst effect.

At low temperature $T\ll T_F$ and within the semiclassical regime, $B \ll B_\text{EQL}$, we can again apply the Mott formula [Eq.~\eqref{eq: Mott formula}]. This calculation gives
\begin{align}
    S_{xy}^{\text{I}} \simeq \frac{\pi^2}{3}  \frac{k_B}{e} \frac{B}{B_1}  \frac{T}{T_F}
    \label{eq: SxyI}
\end{align}
for the regime of $\omega_c(\epsilon = E_F) \tau \ll 1$ (small Hall angle) and 
\begin{align}
    S_{xy}^{\text{II}} \simeq \frac{\pi^2}{3} \frac{k_B}{e} \frac{B_1}{B}  \frac{T}{T_F}
   \label{eq: SxyII}
\end{align}
for the regime of $\omega_c(\epsilon = E_F)  \tau \gg 1$ (large Hall angle). Thus, the Nernst coefficient achieves a peak value ${S_{xy} \sim (k_B/e) (T/T_F)}$ at a field $B \sim B_1$. 

At high temperatures $T\gg T_F$, we must again consider the coexistence of thermally excited electrons and holes. For small enough fields that $\omega_c(\epsilon = k_B T)  \tau \ll 1$, we find
\begin{align}
    S_{xy}^{\text{III}} \simeq \frac{k_B}{e} \frac{B}{B_1} \frac{T_F}{T}.
    \label{eq: SxyIII}
\end{align}
As the magnetic field is increased, this linear increase of $S_{xy}$ with $B$ continues uninterrupted until sufficiently large fields that $\sigma_{xy} \sim \sigma_{xx}$. In other words, 
\begin{align}
    S_{xy}^{\text{IV}} = S_{xy}^\text{III} \simeq \frac{k_B}{e} \frac{B}{B_1} \frac{T_F}{T}.
    \label{eq: SxyIV}
\end{align}
At large enough fields that $\sigma_{xy} \gg \sigma_{xx}$ the value of $S_{xy}$ declines again with $B$ according to
\begin{align}
        S_{xy}^{\text{V}} \simeq (9 \zeta(3))^2 \frac{k_B}{e} \frac{B_1}{B} \left(\frac{T}{T_F}\right)^5.
    \label{eq: SxyV}
\end{align}
These results imply that at $T \gg T_F$ the maximum in $S_{xy}$ is achieved not at $B \sim B_1$; instead, the linear-in-$B$ growth of the Nernst coefficient is continued to a much larger field $B \sim (T/T_F)^3 B_1$. Correspondingly, the maximum value of the Nernst coefficient is not of order $k_B/e$, but is much larger: $S_{xy}^\text{(max)} \sim (k_B/e) (T/T_F)^2$. This large enhancement of the peak Nernst coefficient by magnetic field is reminiscent of a similar effect in conventional semimetals \cite{xiaozhoufeng}.

\section{Circular nodal line}
\label{sec: circular}

\begin{figure}[htb]
    \centering
    \includegraphics[width=\columnwidth]{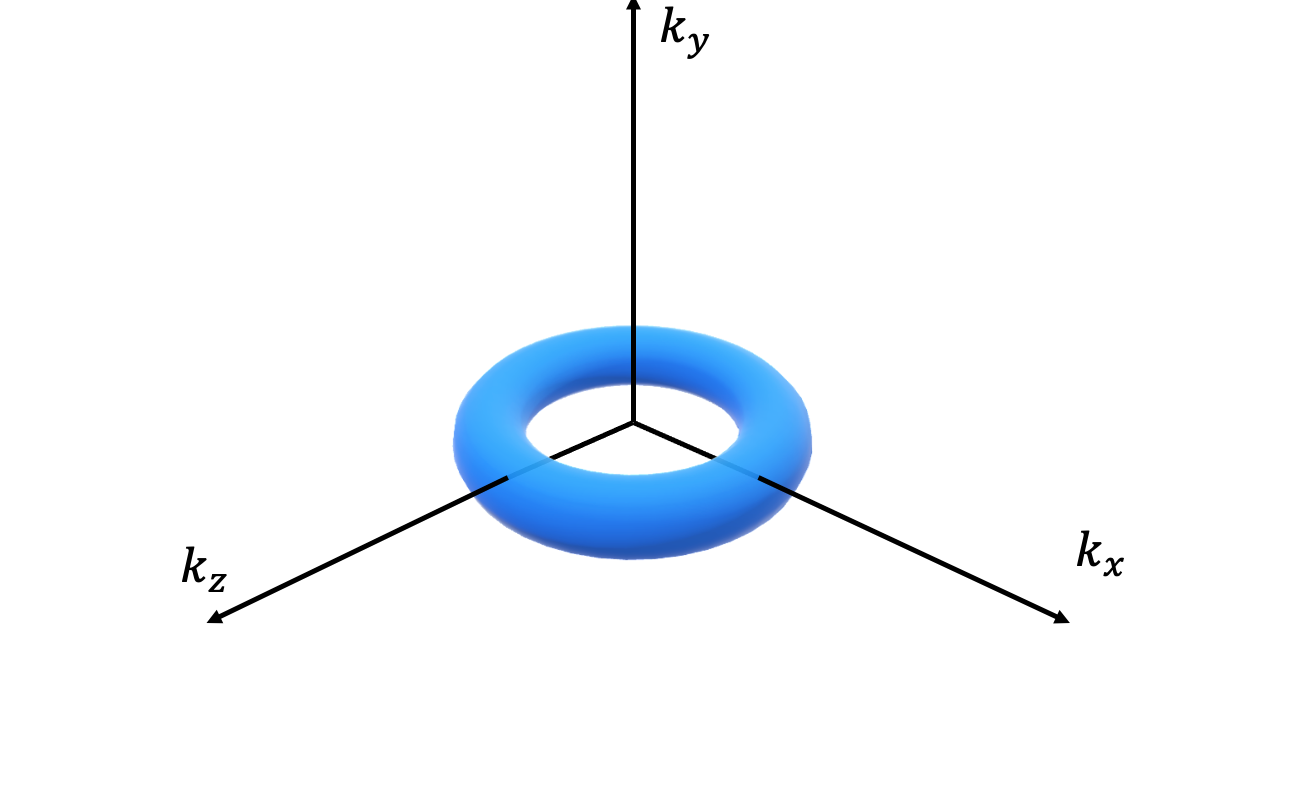}
    \caption{Schematic of the Fermi surface of a NLS with a circular nodal line in the $k_x$ -- $k_z$ plane.}
    \label{fig: circularnodalline}
\end{figure}

In this section we consider the applicability of our results to the case where the conduction and valence bands meet at a circle in momentum space rather than a straight line. Such (approximately) circular nodal lines appear in materials such as ZrSiS \cite{zrsis}, HfSiS \cite{hfsis}, and Ca$_3$P$_2$ \cite{xie_new_2015}. One model Hamiltonian for describing a circular nodal line in the $k_x$ -- $k_z$ plane is \cite{kim_dirac_2015, huh_long-range_2016}
\begin{align}
   H = \frac{\hbar^2}{2m} (k_x^2 + k_z^2 - k_0^2) \sigma_x + \hbar v_F k_y\sigma_y ,
   \label{eq: circular nodal line hamiltonian}
\end{align}
where $k_0$ is the radius of the nodal line in reciprocal space. The corresponding conduction and valence bands have a linear dispersion for momenta close to the nodal line, with a Fermi velocity $v_F$ in the $y$ direction (perpendicular to the nodal line) and a velocity $v_\parallel = \hbar k_0 / m$ in the $x$ and $z$ directions (within the plane of the nodal line). The corresponding Fermi surface at finite $E_F$ is depicted in Fig.~\ref{fig: circularnodalline}. Throughout this section we assume that $E_F, k_B T \ll \hbar^2 k_0^2/m$, so that the Fermi surface takes the shape of a thin torus. 

We note that, in general, the nodal line in real materials is not constrained to reside at a single energy. The ``corrugation" of the nodal line provides an additional energy scale which is not within our description. Thus the results we derive here are applicable when either $E_F$ or $k_B T$ is much larger than the corrugation of the nodal line in energy. We discuss this assumption further in the conclusion.

Let us now consider the behavior of the Seebeck coefficient as a function of magnetic field. We first consider the case where $\mathbf{B}$ is in the $x$-$z$ nodal-line plane; without loss of generality we take $\mathbf{B} = B\mathbf{\hat{z}}$. Since the Lorentz force does not act on currents in the $\mathbf{\hat{z}}$-direction, we do not expect $S_{zz}$ to be modified from its zero-field value, and thus the Seebeck coefficient follows either Eq.~\eqref{eq: SxxI} or \eqref{eq: SxxIII}, depending on whether $T$ or large or small compared to $T_F$. The thermopower $S_{xx}$ and $S_{yy}$ along the directions orthogonal to $\mathbf{B}$, on the other hand, do exhibit a magnetic field enhancement. For $S_{xx}$, the dominant contribution to thermoelectric transport arises from regions of the Fermi surface which have high velocity along the $x$ direction. These are the regions where the nodal ring is nearly parallel to $\mathbf{B}$. Since the Fermi surface is locally cylindrical, the behavior of the thermopower reduces to an analogue of the straight nodal line case considered in Sec.~\ref{sec: straight}, with some ``effective length'' of the nodal line that is of order $k_0$ \cite{syzranov_electron_2017}. Correspondingly, we expect $S_{xx}(B)$ to match the behavior outlined in Fig.~\ref{fig: sxxmap} up to order-$1$ numeric prefactors. For the thermopower $S_{yy}$ along the $y$ direction, the entirety of the nodal line is strongly dispersive along the $y$ direction, but only those parts of the Fermi surface that are nearly parallel to $\mathbf{B}$ experience a significant Lorentz force. For such regions the thermopower is again similar to what is shown in Fig.~\ref{fig: sxxmap}. So we arrive at the conclusion that both $S_{xx}(B)$ and $S_{yy}(B)$ are equivalent to the case of the straight nodal line up to numeric prefactors. (We are neglecting in our discussion the effects of weak localization, which can provide a correction to the conductivity tensor at low magnetic field that is strongly dependent on the field direction \cite{syzranov_electron_2017}.)

\begin{figure}
    \centering
    \includegraphics[width=\columnwidth]{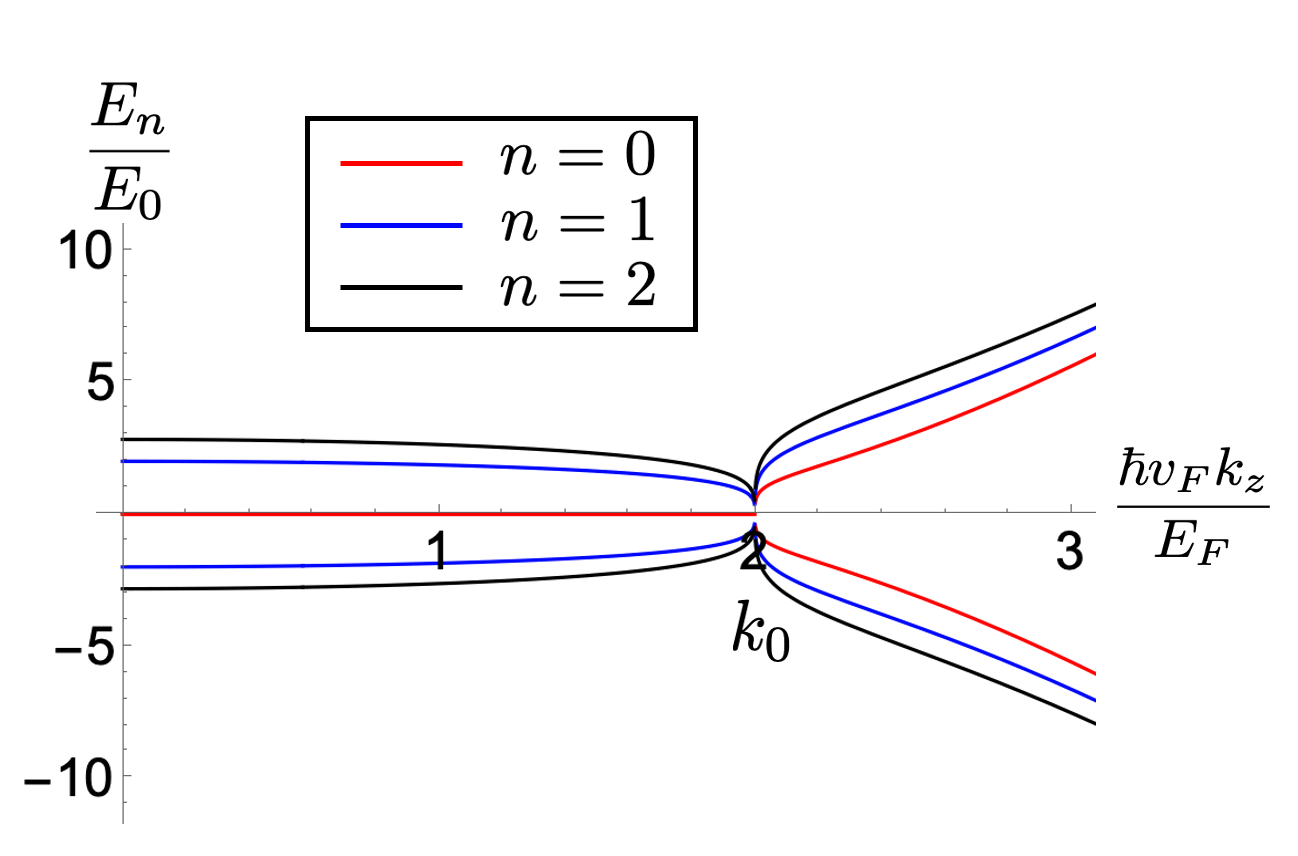}
    \caption{Dispersion relation for the lowest three Landau levels of a NLS with a circular nodal line in the $k_x$ -- $k_z$ plane, plotted as a function of the momentum $k_z$ along the field direction [using Eqs.~(\ref{eq: Landau levels pz < p0}) and (\ref{eq: Landau levels pz > p0})]. This example uses $k_0 = 2E_F/(\hbar v_F)$. The unit of energy on the y-axis $E_0 = \sqrt{\hbar^2 e B v_F k_0/m} = \hbar (v_F v_\parallel)^{1/2} / l_B$. Notice that the zeroth Landau level (red curve) is electron-hole degenerate for $|k_z| < k_0$. }
    \label{fig: landaulevelscircularnodalline}
\end{figure}

Let us consider in detail the behavior of the Seebeck coefficient in the EQL (regime VI). For the case of a straight nodal line, the large, linear-in-$B$ and temperature-independent value of $S_{xx}$ [Eq.~\eqref{eq: SxxVI}] arises from the existence of a zero-energy Landau level that is shared by conduction and valence band states and that does not disperse along the field direction. In fact, the circular nodal line also exhibits a zero-energy Landau level that is dispersionless for $k_z$ in the range $-k_0 < k_z < k_0$ (see Fig.~\ref{fig: landaulevelscircularnodalline}); this result was derived in Refs.\ \onlinecite{NLSM_Burkov, rhim_landau_2015, yan_nodal_2017}. We also provide a detailed derivation of the Landau level spectrum in Appendix \ref{appendix: circular nodal line}. The dispersionless portion of the zero-energy Landau level enables a Seebeck coefficient enhancement that is very similar to that of Eq.~\eqref{eq: SxxVI}.

Quantitatively, we can estimate the value of $S_{xx}(B)$ at $B \gg B_\text{EQL}$ by assuming that carriers half-populate the flat portion of the $n=0$ Landau level. At temperatures which are low compared to $\sqrt{\hbar^2 v_F k_0 e B/m}$ (the spacing between the zeroth and first Landau levels at small $k_z$), the configurational entropy of carriers within this flat portion of the zeroth Landau level provides the primary contribution to the entropy. The degeneracy of the flat band is $N_{\textrm{flat}} = gk_0/(\pi l_B^2)$ per volume, where $l_B = \sqrt{\hbar/eB}$ is the magnetic length. The flat band is half-filled and therefore the entropy density is 
\begin{align}
\label{eq: entropy circular nodal line EQL}
    n_s = k_B \ln{{ N_{\textrm{flat}} \choose \frac{N_{\textrm{flat}}}{2}}} \simeq k_B N_{\textrm{flat}} \ln{2} = k_B \frac{g k_0}{\pi l_B^2} \ln{2}.
\end{align}
The charge density is given by
\begin{align}
    \rho_e = e g \frac{\pi k_{F\perp} k_{F\parallel}\cdot (2 \pi k_0)}{(2 \pi)^3} = \frac{g m E_F^2 e}{4\pi \hbar^3 v_F}.
    \label{eq: charge density circular nodal line EQL}
\end{align}
Here, the numerator in the first equality of Eq.~\eqref{eq: charge density circular nodal line EQL} represents the volume of the toroidal Fermi surface, with $k_{F\perp}$ and $k_{F\parallel}$ being the Fermi momentum in the directions perpendicular ($y$-direction) and parallel ($x$-$z$ plane) to the circular nodal line, respectively. Note that the toroidal Fermi surface need not have circular cross-section. In the second equality, we use the relations $E_F = \hbar v_F k_{F\perp} = \hbar v_{\parallel}  k_{F\parallel}$, where $E_F$ is the energy measured relative to the nodal line. 
Thus, we find
\begin{align}
    S_{xx}^\text{VI} \simeq  4 \ln{2} \frac{k_B}{e} \frac{e\hbar^2 v_F k_0}{m E_F^2} B.
\end{align}

As argued above, we recover the large, linear-in-$B$ and temperature-independent behavior of the Seebeck coefficient obtained for the case of a straight nodal line. In fact, one can obtain this result directly from Eq.~\eqref{eq: SxxVI} by making the simple replacement $v_F^2 \rightarrow v_\parallel v_F$ and inserting an additional factor of $2$ associated with the fact that occupied states range from $-k_0$ to $k_0$.

Finally, let us consider the case where $\mathbf{B} = B\mathbf{\hat{y}}$ is applied out of the nodal-line plane. At zero magnetic field, the thermopower is identical to the case of the straight nodal line, since the Fermi surface is everywhere locally cylindrical. Significant magnetic field effects appear only when $\omega_c \tau$ becomes order-$1$ or larger, where in this case the value of $\omega_c$ depends on the mass associated with cyclotron motion in the $x$-$z$ plane. For such motion the effective mass is of order $m$ [where the parameter $m$ is defined by the Hamiltonian, Eq.~\eqref{eq: circular nodal line hamiltonian}], rather than the much smaller effective mass associated with ``out of plane'' cyclotron motion: $m_\perp \equiv \hbar k_F/v_F$. 
When $m_\perp/m \sim \hbar \sqrt{n_e}/(v_F m \sqrt{k_0}) \ll 1$, i.e.\ for sufficiently large $k_0$ or low electron concentration, the strength of an out-of-plane $B$-field necessary to produce significant magnetic field effects is much larger than for the in-plane case. For example, a NLS with charge density $n = 10^{18}$\,cm$^{-3}$, Fermi velocity $v_F \sim 10^5 $ m/s and nodal line radius $k_0 \sim 0.1$\,\AA$^{-1}$ would have $m_\perp/m \sim 0.03$, so that producing strong magnetic effects by an out-of-plane magnetic field would require a thirty times larger field than for the case of an in-plane magnetic field. We therefore leave the case of out-of-plane field to be explored by future work.

\section{Summary and outlook}
\label{sec: conclusion}

We have studied the behavior of the thermoelectric coefficients of nodal line semimetals across all regimes of temperature and magnetic field. Our results broadly apply regardless of the shape of the nodal line, as discussed in Sec.~\ref{sec: circular}. We find that magnetic field produces a large enhancement of both the Seebeck and Nernst coefficients, as depicted in Figs.~\ref{fig: sxxmap} -- \ref{fig: sxy map}, so that both can be driven well above  $k_B/e$. 

It should be emphasized that it is usually difficult in real materials to achieve  a thermopower that is parametrically enhanced above the natural unit $k_B/e \approx 86$\,$\mu$V/K within the metallic state, due to the fundamental tradeoff between having low electronic entropy at low temperature and having cancellation between electron and hole contributions to thermal transport at high temperature. But a strong magnetic field is able to circumvent this limitation by enabling a transverse mechanism of thermoelectric transport in which electrons and holes carry heat in parallel via the $\mathbf{E} \times \mathbf{B}$ drift \cite{Skinnereaat2621}. Some preliminary evidence for this strong magnetic field enhancement has been seen in the NLS Mg\textsubscript{3}Bi\textsubscript{2} \cite{NL_Feng_Mg3Bi2}.

Our most dramatic finding is that at sufficiently large magnetic field that only the lowest Landau level is occupied, i.e.\ in the extreme quantum limit, a NLS exhibits a large, linear-in-$B$ and temperature-independent Seebeck coefficient, as given by Eq.~\eqref{eq: SxxVI}. This large thermopower arises fundamentally from the huge entropy associated with a half-filled and electron-hole degenerate lowest Landau level, which is a hallmark of topological semimetals. As we discussed in Sec.~\ref{sec: circular}, this feature exists for both straight and circular nodal lines. Given the simplicity of this result, it is worth writing Eq.~\eqref{eq: SxxVI} in experimental units:
\begin{align}
    S_{xx}^\text{VI} \simeq 289 \frac{\mu\textrm{V}}{\textrm{K}} \times \frac{(k_0 \, [\textrm{in \AA}^{-1}])(B \, [\textrm{in T}])}{n_e \, [\textrm{in $10^{18}$ cm}^{-3}]}.
\end{align}
In this expression, $2 \pi k_0$ corresponds to the length of the nodal line. Notice that even under very realistic experimental conditions, such as $k_0 \sim 0.1$\,\AA$^{-1}$, $B = 10$\,T, and $n_e = 10^{18}$\,cm$^{-3}$, it becomes possible to attain a large thermopower $S_{xx} \approx 300$\,$\mu$V/K even at very low temperature. Under more optimistic conditions, where the carrier concentration is reduced significantly below $10^{18}$\,cm$^{-3}$, it may be possible to realize thermopower on the order of mV/K at low temperature. 

It should be noted that for most practical applications the thermodynamic efficiency is determined not by the Seebeck coefficient alone but by the thermoelectric figure of merit $zT = S_{xx}^2 T/\rho \kappa$, which depends on the electrical resistivity $\rho$ and the thermal conductivity $\kappa$. (In the presence of a magnetic field both $\rho$ and $\kappa$ are tensors, but this expression for the figure of merit remains accurate with the substitutions $\rho \rightarrow \rho_{xx}$ and $\kappa \rightarrow \kappa_{xx}$  so long as $\kappa_{xy} \ll \kappa_{xx}$ \cite{Skinnereaat2621}.) 
In materials with low Fermi energy, $\kappa$ is dominated by phonons at all but extremely small temperatures, while the resistivity $\rho_{xx}$ may be subject to strong and nontrivial magnetoresistance effects.  Calculating these quantities is therefore beyond the scope of this paper.  However, we note that within the EQL (regime VI) one can write the figure of merit as
\begin{align}
    zT \simeq 0.36 \frac{ (B/B_\text{EQL})^2 }{(\kappa_{xx}/T \, [\textrm{in W}\textrm{m}^{-1}\textrm{K}^{-2}])(\rho_{xx} \, [\textrm{in $\mu \Omega$cm}])}.
\end{align}
The quantities in the denominator of this expression can realistically be of order unity (e.g., in ZrSiS \cite{singha_large_2017, husain_electron_2020}). Achieving large $zT$ at low temperature therefore seems plausible if the EQL can be achieved, or in other words if the magnetic field is strong enough and the level of unintentional doping is sufficiently low. Of course, practical applications will also require reducing the field scale $B_\textrm{EQL}$ to the level that can be achieved by a permanent magnet (below $1 - 2$ T), so that additional power input is not required to generate the magnetic field.

So far we have assumed the nodal line to be flat in energy, meaning that we neglect variations in energy along the nodal line (``corrugations'' of the nodal line). In general there is no symmetry that constrains a nodal line to reside at constant energy (e.g., the function $F(p_z)$ in Eq.~\eqref{eq: Hstraight} need not be zero)   , and any such variation introduces an additional energy scale that competes with the magnetic field effects we are discussing. Specifically, when the Fermi energy is small compared to the corrugation energy scale, the toroidal Fermi surface depicted in Fig.~\ref{fig: circularnodalline} breaks up into contiguous electron and hole pockets. Such an effect is apparently prominent in \mbox{ZrSiS}, for example \cite{zrsis, neupane_observation_2016}, and it puts a lower limit on the achievable carrier density that is apparently of order $n_e \sim 10^{19}$\,cm$^{-3}$ \cite{singha_large_2017, matusiak_thermoelectric_2017}. Practical efforts to achieve low-temperature thermoelectrics using NLSs may therefore find it fruitful to identify materials with a nearly flat nodal line.

Finally, we mention that throughout this paper we have worked within a noninteracting band picture, ignoring the effects of electron-electron interactions. Such a picture is not generally applicable in a flat band at very low temperature, since the exchange interaction tends to split a degenerate band into spin- or orbital-polarized subbands. For example, two-dimensional graphene has a zero energy Landau level that is nominally four-fold degenerate, but at liquid helium temperatures and high fields this Landau level is split by the exchange interaction into spin- and valley-polarized subbands \cite{zhao_symmetry_2010}. This exchange splitting limits the entropy of the electron system and bounds the measured Seebeck coefficient at just a few times $k_B/e$ \cite{ghahari_kermani_interaction_2014}. Whether such exchange splitting is relevant for three-dimensional NLSs remains to be seen, but if it is relevant then its primary effect will be to limit the applicability of our results to temperatures above the exchange splitting scale.

\acknowledgments 

The authors thank Mingda Li for helpful discussions.
This work was primarily supported by the Center for Emergent Materials, an NSF-funded MRSEC, under Grant No. DMR-2011876. G.B.\ is supported by the U.S. Department of Energy (DOE) Established Program to Stimulate Competitive Research (EPSCoR) Grant No.\ DE-SC0024284.

\appendix


\section{Comments on the chemical potential}
\label{app: constantmu}



\begin{figure} [h!]
    \centering
    \includegraphics[width = \columnwidth]{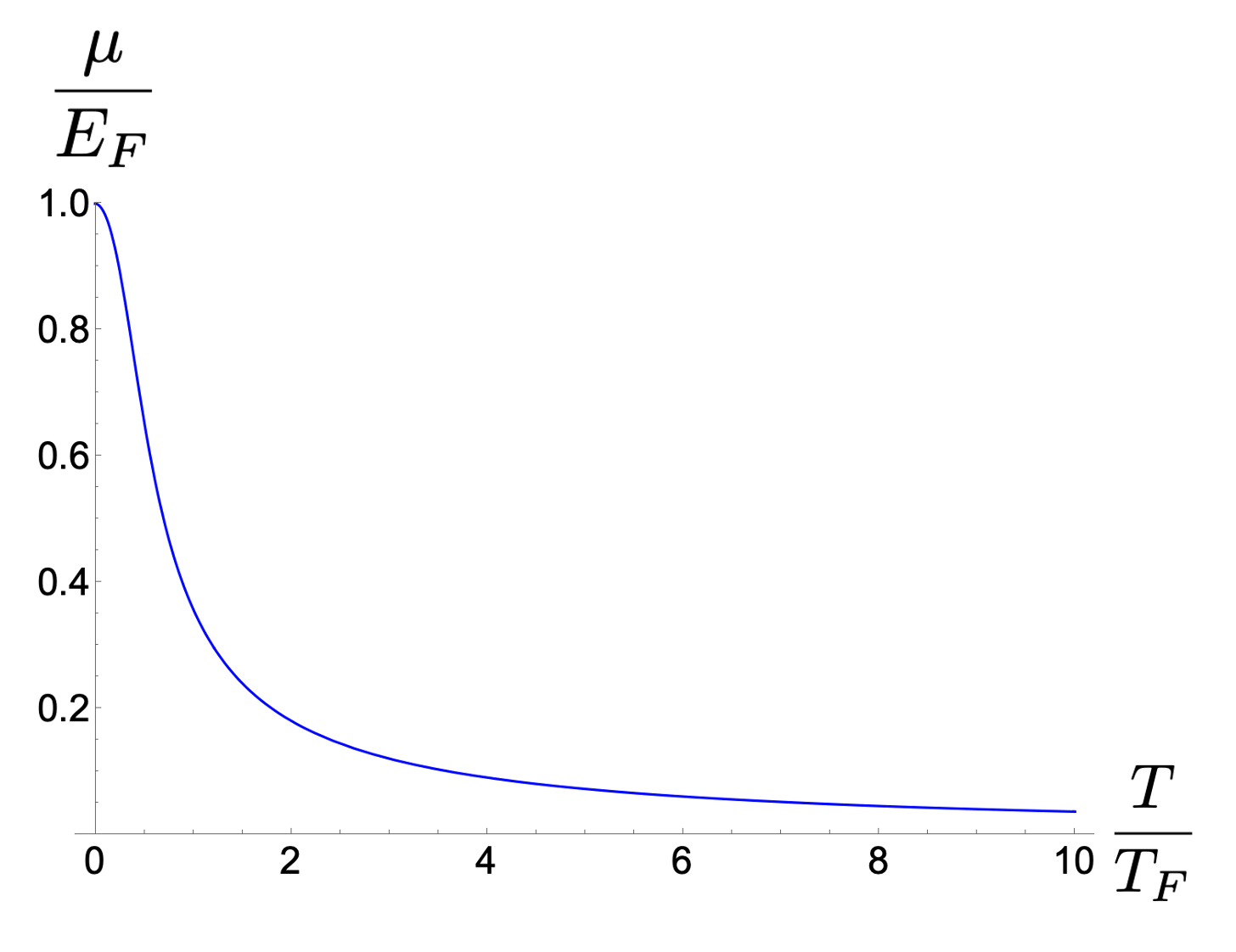}
    \caption{Chemical potential $\mu$ as a function of temperature $T$ for the case of fixed charge density in the nodal-line bands. }
    \label{fig: mu vs T}
\end{figure}

Throughout the main text we have assumed that the charge density $n_e - n_h$ of carriers within the nodal-line bands remain constant as a function of magnetic field and temperature. This assumption is natural if there are no other electron bands within $\sim k_BT$ of the Fermi level. Using this assumption we can calculate the dependence of the chemical potential on temperature using Eq.~\eqref{eq: mu vs T}, which gives the result shown in Fig.~\ref{fig: mu vs T}. At low temperature the value of $\mu$ approaches $E_F$, while at high temperatures $T \gg E_F$ we have $\mu \simeq E_F^2/(4 k_BT \ln{2})$.

On the other hand, if there are other bands with high density of states intersecting the chemical potential, then these bands can have the effect of pinning the value of $\mu$ so that it does not evolve (or evolves more weakly) as a function of temperature. Such a situation apparently arises in the Dirac semimetal ZrTe$_5$ \cite{Zhang2020} and in the Weyl semimetal NbP \cite{scott_doping_2023}, for example. Thus, it is worth briefly considering how our results are modified in the case where $\mu$ is constant as a function of temperature.

\begin{figure} [h!]
    \centering
    \includegraphics[width =\columnwidth]{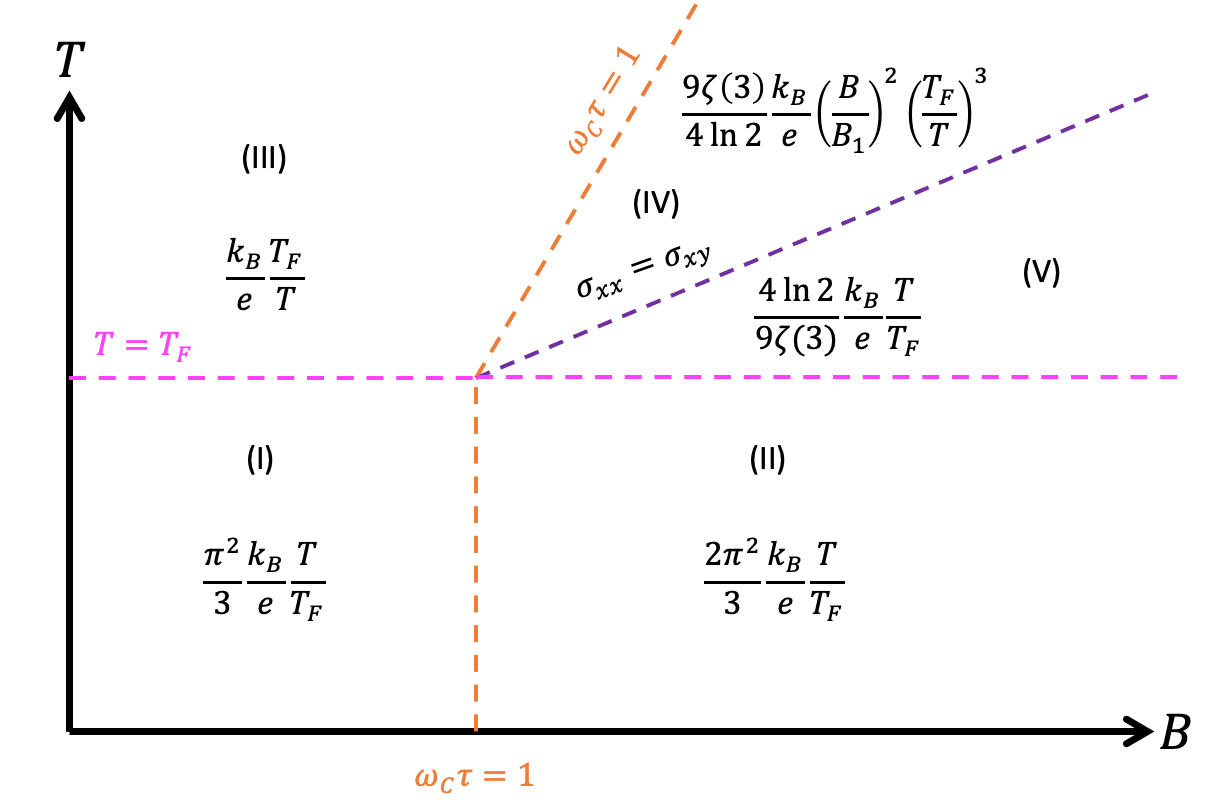}
    \caption{A schematic plot of the Seebeck coefficient $S_{xx}$ against $B$ and $T$ for a NLS with a straight nodal line in various asymptotic regimes, assuming a constant chemical potential $\mu$ as a function of $T$. The two axes are depicted on a logarithmic scale and the various regimes of $T$ and $B$ are the same as in Fig.~\ref{fig: sxxmap}. $B_1$ is as defined in Eq.~\eqref{eq: B1 definition}.}
    \label{fig:seebeckconstantmu}
\end{figure}

\begin{figure} [h!]
    \centering
    \includegraphics[width = \columnwidth]{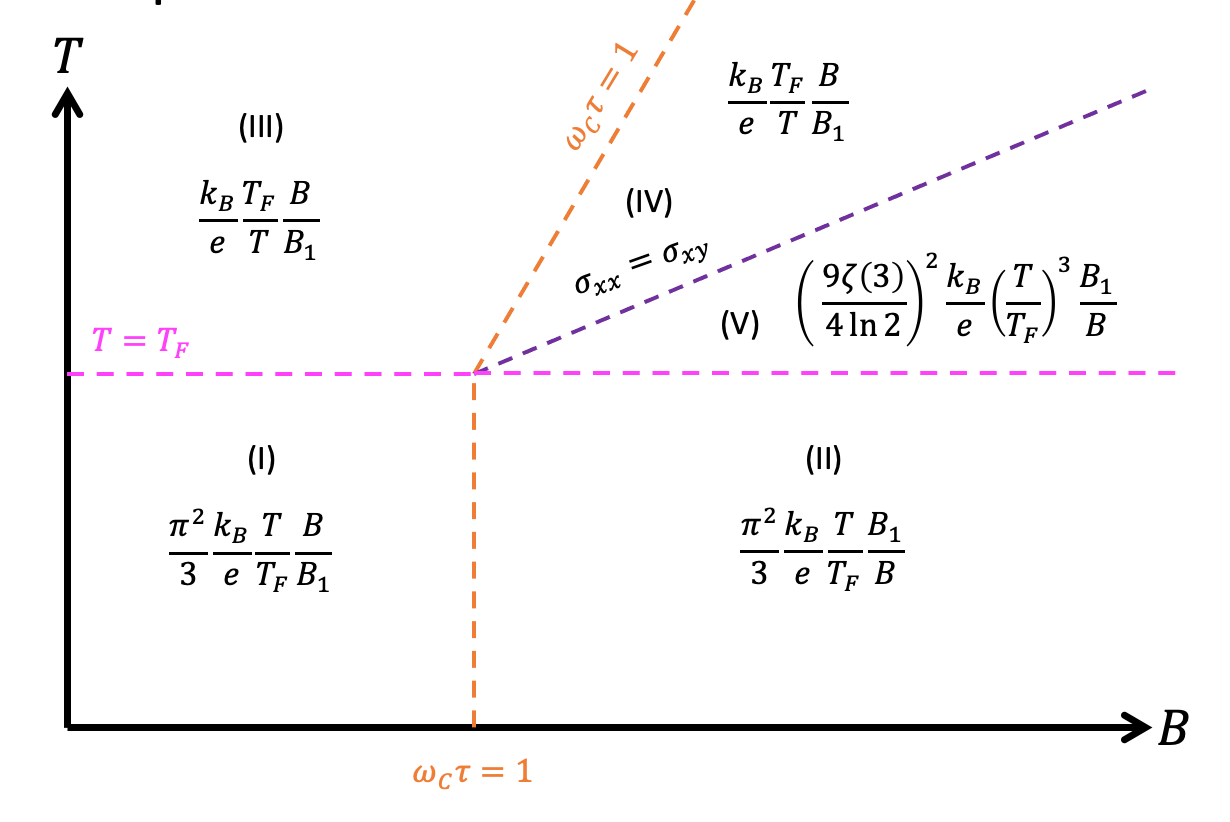}
    \caption{A schematic plot of the Seebeck coefficient $S_{xy}$ against $B$ and $T$ for a NLS with a straight nodal line in various asymptotic regimes, assuming a constant chemical potential $\mu$ as a function of $T$. The two axes are depicted on a logarithmic scale and the various regimes of $T$ and $B$ are the same as in Fig.~\ref{fig: sxy map} $B_1$ is as defined in Eq. (\ref{eq: B1 definition}).}
    \label{fig:sxyconstantmu}
\end{figure}

Figures \ref{fig:seebeckconstantmu} and \ref{fig:sxyconstantmu} show the corresponding values of the Seebeck and Nernst coefficients for the case where $\mu$ is constant as a function of temperature. In general the expressions are unmodified at low temperatures $T \ll T_F$, since in this case the chemical potential $\mu$ approaches $E_F$ in any situation. The value of $S_{xx}$ is also unchanged in the EQL (regime VI, not shown in Figs.~\ref{fig:seebeckconstantmu} and \ref{fig:sxyconstantmu}). At higher temperatures $T \gg T_F$ the power-law dependence on temperature is modified due to the chemical potential not sinking down toward the nodal line at higher temperatures, although the qualitative dependence on $T$ remains the same in all regimes. The power-law dependence of both $S_{xx}$ and $S_{xy}$ to $B$ is unchanged in all regimes

\section{Landau levels of a circular nodal line}
\label{appendix: circular nodal line}

The Hamiltonian for electron bands near a circular nodal line is given by Eq. (\ref{eq: circular nodal line hamiltonian}).
The first term describes the circular nodal line in the momentum space, with the radius $k_0 $. The second term describes the linear dispersion in the $k_z$ direction. $\sigma_x$ and $\sigma_y$ are the Pauli matrices.

We assume the magnetic field to be oriented in the plane parallel to the plane of the nodal line, and  in the $z$ direction. We can introduce this field using a Peierls substitution $p_y \rightarrow \hbar k_y - eBx $, 
so that the Hamiltonian can be written as:
\begin{align}
   H = \frac{\hbar^2}{2m} (k_z^2 - k_0^2 - \partial_x^2) \sigma_x + v_F(\hbar k_y - eBx)\sigma_y .
   \label{eq: Hpauli}
\end{align}

\subsection{Zero Energy Modes}
We begin by calculating the eigenstates with zero energy. By substitution, and solving the Schrodinger equation, we find that the possible solutions exist only when $p_x < p_0$. Multiplying out the matrix equation \eqref{eq: Hpauli} for a wave function $(\psi_1, \psi_2)^\textrm{T}$, we arrive at
\begin{align}
   \left[ \frac{1}{2m}(p_z^2 - p_0^2 - \hbar^2 \partial_x^2 ) + i v_F ( p_y - eBx)\right]\psi_1 &= 0
\\
   \left[ \frac{1}{2m}(p_z^2 - p_0^2 - \hbar^2 \partial_x^2 ) - iv_F (p_y - eBx)\right]\psi_2 &= 0.
\end{align}

To solve for $\psi_2$ we rewrite the equation as:
\begin{align}
   \hbar^2 \partial_x^2 \psi_2 = 2miv_F eB \left( x - \frac{p_y}{eB} + \frac{p_z^2 - p_0^2}{2imv_F eB} \right) \psi_2.
\end{align}
Using the new variable
\begin{align}
    x' &= x - \frac{p_y}{eB} + \frac{p_z^2 - p_0^2}{2imv_F eB}
\end{align}
this equation becomes
\begin{align}
   \hbar^2 \frac{\partial^2 \psi_2}{\partial x'^2} &= 2miv_F eB x' \psi_2.
\end{align}

Introducing another new variable $z = (2miv_F eB/\hbar^2)^{1/3} x'$, the equation turns into the Airy equation:
\begin{align}
    \frac{\partial^2 \psi_2}{\partial z^2} = z \psi_2.
\end{align}
The solution to this equation is the Airy function 
\begin{align}
    \psi_2 = \textrm{Ai}(z).
\end{align}


Coming back to the original variable,
\begin{align}
    \psi_2 = \textrm{Ai} \left( \left(\frac{2imv_F eB}{\hbar^2}\right)^{1/3} \left( x - \frac{p_y}{eB} + \frac{p_z^2 - p_0^2}{2imv_FeB} \right) \right) 
\end{align}

Using the asymptotic expression for Airy functions as the argument $z \rightarrow \infty $,
\begin{align}
    \textrm{Ai} (z) \simeq \frac{\exp \left[-\frac{2}{3}z^{3/2} \right]}{2 \sqrt{\pi} z^{1/4}}
\end{align}
we get
%
\begin{align}
    \psi_2
    \simeq \frac{\exp\left[-\frac{2}{3}\left(2imv_F eB/\hbar^2\right)^{1/2}\left( x - \frac{p_y}{eB} + \frac{p_z^2 - p_0^2}{2imv_FeB} \right)^{3/2} \right] }{2 \sqrt{\pi} \left(2imv_F eB/\hbar^2\right)^{1/12}\left( x - \frac{p_y}{eB} + \frac{p_z^2 - p_0^2}{2imv_FeB} \right)^{1/4}}.
\end{align}
In the limit $x \rightarrow \infty $ this expression decays exponentially. 

Now, we would study the behavior of the wave function when the $x$ variable changes from $\infty $ to $-\infty$, and therefore when the argument $z$ of the Airy function changes from $e^{i\pi/6} \infty$ to $e^{-5i\pi/6} \infty$ or $e^{7i\pi/6} \infty$. For $p_z > p_0$, in the complex plane $y$ lies below zero; to obtain continuous expression we assume at $x \rightarrow - \infty$, $z \rightarrow e^{-5i\pi/6} \infty$ and therefore $z^{3/2} \rightarrow e^{-5i\pi/4} \infty $. For $p_z < p_0$, in the complex plane $x$ lies below zero; to obtain continuous expression we assume at $x \rightarrow - \infty$, $z \rightarrow e^{\frac{7i\pi}{6}} \infty$ and therefore $z^{3/2} \rightarrow e^{7i\pi/4} \infty $. Therefore at $x \rightarrow - \infty$, $\psi_2$ only decays at $p_z < p_0$

The calculation for $\psi_1$ is very similar to the calculation for $\psi_2$. The equation is:
\begin{align}
   \hbar^2 \partial_x^2 \psi_1 = -2miv_F eB \left( x - \frac{p_y}{eB} + \frac{p_z^2 - p_0^2}{-2imv_F eB} \right) \psi_1.
\end{align}
Similarly, we declare a new variable
\begin{align}
    x' = x - \frac{p_y}{eB} + \frac{p_z^2 - p_0^2}{-2imv_F eB},
\end{align}
and the equation becomes 
\begin{align}
    \hbar^2 \frac{\partial^2 \psi_1}{\partial x'^2} = - 2miv_F eB x' \psi_1.
\end{align}
Introducing a new variable $z = \left(\frac{-2miv_F eB}{\hbar^2}\right)^{\frac{1}{3}} x'$, the equation turns into Airy equation:
\begin{align}
    \frac{\partial^2 \psi_1}{\partial z^2} = z \psi_1.
\end{align}
The solution of this equation is 
\begin{align}
    \psi_1 = \textrm{Ai} (z) \sim \frac{e^{-\frac{2}{3}z^{3/2}}}{2 \sqrt{\pi} z^{1/4}}.
\end{align}

At the limit $x \rightarrow \infty $, $z^{3/2} \sim e^{-i\pi/4}$, this expression decays exponentially. At the limit $ x \rightarrow -\infty$, $z^{3/2} \sim e^{-7 i \pi/4} \infty$ for $p_z < p_0$ and $z^{3/2} \sim e^{5i\pi/4} \infty$ at $p_z > p_0$. So we conclude that zero modes exist only at $p_z < p_0$.

\subsection{Solutions with non-zero energies}

\subsubsection{Preliminary Steps}

In order to calculate the non-zero energy eigenstates, we first write the wave function in terms of its Fourier transform:
\begin{align}
    \psi_{1(2)} (p_z,p_y,x) = \int \frac{dp_x}{2\pi 
\hbar} e^{ip_x x/\hbar} \psi_{1(2)} (p_x,p_y,p_z).
\end{align}
The $\hat{x}$ operator transforms as $\hat{x} \rightarrow i\hbar \frac{\partial}{\partial p_x}$. The Hamiltonian in momentum representation can be written as:
\begin{align}
   H = \frac{1}{2m} (p_x^2 + p_z^2 - p_0^2) \sigma_x + v_F \left(p_y - ieB\hbar \frac{\partial}{\partial p_x} \right)\sigma_y .
\end{align}
We introduce a change of variables
\begin{align}
    \psi_{1(2)} (p) = e^{-ip_x p_y/eB\hbar} \phi_{1(2)},
\end{align}
so that the eigenvalue problem $H\psi = E\psi$ looks like 
\begin{align}
    \frac{E}{v_F eB} \begin{pmatrix} \phi_1 \\ \phi_2 \end{pmatrix} &= \begin{pmatrix} 0 & A^{\dagger} \\ A & 0 \end{pmatrix}  \begin{pmatrix} \phi_1 \\ \phi_2 \end{pmatrix}.
\end{align}
Here, $A$ and $A^{\dagger}$ are defined as
\begin{align}
A &= \hbar \frac{\partial}{\partial p_x} + \frac{p_x^2 + p_z^2 - p_0^2}{2mv_F eB} 
\\
 A^{\dagger} &= -\hbar \frac{\partial}{\partial p_x} + \frac{p_x^2 + p_z^2 - p_0^2}{2mv_F eB}. 
 \end{align}
The corresponding eigenvalues and eigenfunctions can be solved using the following equations:
\begin{align}
     \left( \frac{E}{v_F eB} \right)^2 \phi_1 &= A^{\dagger} A \phi_1 
     \\
 \left( \frac{E}{v_F eB} \right)^2 \phi_2 &= AA^{\dagger}  \phi_2 .
\end{align}
The explicit forms of the operators on the RHS are:
\begin{align}
     A^{\dagger} A &= -\hbar^2 \frac{\partial^2}{\partial p_x^2} + \frac{(p_x^2 +p_z^2 -p_0^2)^2}{(2mv_F eB)^2} - \frac{\hbar p_x}{mv_F eB} 
     \\
 AA^{\dagger} &= -\hbar^2 \frac{\partial^2}{\partial p_x^2} + \frac{(p_x^2 +p_z^2 -p_0^2)^2}{(2mv_F eB)^2} + \frac{\hbar p_x}{mv_F eB} .
\end{align}

In the remainder of this subsection we derive solutions to these eigenvalue equations for the separate cases where the momentum $|p_z| < p_0$ and $|p_z| > p_0$.

\subsubsection{$ |p_z| < p_0 $}

We now evaluate the energy levels for the case where $ |p_z| < p_0 $. 
%
The Schrodinger equation in this case is given by
\begin{align}
    \left( \frac{E}{v_F eB} \right)^2 \phi_2 & = \left[ -\hbar^2 \frac{\partial^2}{\partial p_x^2} + V(p_x) \right] \phi_2,
\end{align}
where
\begin{align}
V(p_x) & = \frac{(p_x^2 + p_z^2 - p_0^2)^2}{(2mv_F eB)^2} + \frac{\hbar p_x}{mv_F eB} 
\end{align}
plays the role of a ``potential'' in momentum space. The potential $V(p_x)$ has the form of a ``Mexican hat'' potential with two degenerate minima as a function of $p_x$ located at 
\begin{align}
    p_x = \pm \sqrt{p_0^2 - p_z^2} - \frac{\hbar mv_F eB}{2 (p_0^2 - p_z^2)} . 
    \label{eq: px approx}
\end{align}
Expanding the potential around these minima and using the limit $ 4m \hbar v_F eB \ll (p_z^2 - p_0^2)^{3/2}$, the potential can be written as
\begin{align}
    V(p_x) \approx \pm \frac{\sqrt{p_0^2 - p_z^2}}{\hbar^2 mv_FeB} + \frac{ p_0^2 - p_z^2}{(\hbar mv_FeB)^2} (\delta p_x)^2,
    \label{eq: Vpx}
\end{align}
where $\delta p_x$ is given by
\begin{align}
    \delta p_x = p_x - \left[ \pm \sqrt{p_0^2 - p_z^2} - \frac{\hbar mv_F eB}{2 (p_0^2 - p_z^2)} \right].
\end{align}
We justify our assumption of small $\delta p_x$ below.

Equation \eqref{eq: Vpx} implies that the Schrodinger equation becomes that of a one-dimensional harmonic oscillator, which gives us the following energy eigenvalues
\begin{align}
    E = \pm (p_0^2 - p_z^2)^{1/4} \sqrt{\frac{2\hbar v_F eB(n+1)}{m}} ,
\end{align}
for the minima denoted by '$+$' and
\begin{align}
    E = \pm (p_0^2 - p_z^2)^{1/4} \sqrt{\frac{2\hbar v_F eBn}{m}}
    \label{eq: Landau levels pz < p0}
\end{align}
for the minima denoted by '$-$'. In both cases the integer $n = 0, 1, 2, ... \dots$. This latter equation implies the existence of a zero-energy Landau level at $p_z < p_0$.


\subsubsection{$|p_z| > p_0 $}

We now calculate the energy values for the case $|p_z| > p_0$. If we assume again that $\delta p_z$ is small enough that $ 4m \hbar v_F eB \ll (p_z^2 - p_0^2)^{3/2}$, then we arrive at a parabolic potential $V(p_x)$ with only one minimum located at 
\begin{align}
    p_x = - \frac{\hbar mv_F eB}{p_z^2 - p_0^2} . 
\end{align}
Expanding $V(p_x)$ around this minima by $\Delta p_x$, the Schrodinger equation again becomes that of a harmonic oscillator:
\begin{align}
    (2mE)^2 \phi_2 \simeq & \left[ -4 (\hbar mv_F eB)^2 \frac{\partial^2}{\partial p_x^2} + \ldots \right.  \\
    &\quad  \left. (p_z^2 - p_0^2)^2 + 2(p_z^2 - p_0^2) \left(\delta p_x\right)^2 \right] \phi_2  . \nonumber
\end{align}    
The corresponding energy eigenvalues are
\begin{align}
    E = \pm \frac{1}{2m} \sqrt{(p_z^2 - p_0^2)^2 + 2\sqrt{2}\hbar mv_F eB\sqrt{p_z^2 - p_0^2} (2n +1)}.
    \label{eq: Landau levels pz > p0} 
\end{align}

The results from Eqs.~\eqref{eq: Landau levels pz < p0} and \eqref{eq: Landau levels pz > p0} are plotted in Fig.~\ref{fig: landaulevelscircularnodalline} of the main text for $n = 0, 1, 2$.

\bibliography{reference.bib}

\end{document}